\documentclass[aps,prd,amsmath,showpacs,superscriptaddress,nofootinbib,twocolumn]{revtex4}
\usepackage{graphicx,epsfig}
\usepackage{bm}
\def\roughly#1{\mathrel{\raise.3ex\hbox{$#1$\kern-.75em%
\lower1ex\hbox{$\sim$}}}}

\begin{document}

\title{Influence of pions and hyperons on stellar black hole formation}

\author{Bruno Peres}
\affiliation{Laboratoire Univers et Th\'eories (LUTH), Observatoire de
  Paris/CNRS/Universit\'e Paris Diderot, 5 place Jules Janssen, 92195 Meudon, France}

\author{Micaela Oertel}
\affiliation{Laboratoire Univers et Th\'eories (LUTH), Observatoire de
  Paris/CNRS/Universit\'e Paris Diderot, 5 place Jules Janssen, 92195 Meudon, France}

\author{J\'er\^ome Novak}
\affiliation{Laboratoire Univers et Th\'eories (LUTH), Observatoire de
  Paris/CNRS/Universit\'e Paris Diderot, 5 place Jules Janssen, 92195 Meudon, France}

\date{December, $21^{\rm st}$ 2012}

%%%%%%%%%%%%%%%%%%%%%%%%%%%%%%%%%%%%%%%%%%%%%%%%%%%%%%%%

\begin{abstract}
  We present numerical simulations of stellar core-collapse with spherically
  symmetric, general relativistic hydrodynamics up to black hole
  formation. Using the \textsc{CoCoNuT} code, with a newly developed grey
  leakage scheme for the neutrino treatment, we investigate the effects of
  including pions and $\Lambda$-hyperons into the equation
  of state at high densities and temperatures on the black hole formation
  process. Results show  non-negligible differences between the
  models with reference equation of state without any additional particles and
  models with the extended ones. For the latter, the maximum masses
  supported by the proto-neutron star are smaller and the collapse to a black
  hole occurs earlier. A phase transition to hyperonic matter is observed when
  the progenitor allows for a high enough accretion rate onto the
  proto-neutron star. Rough estimates of neutrino luminosity from these
  collapses are given, too.
\end{abstract}

%%%%%%%%%%%%%%%%%%%%%%%%%%%%%%%%%%%%%%%%%%%%%%%%%%%%%%%%


\maketitle

%%%%%%%%%%%%%%%%%%%%%%%%%%%%%%%%%%%%%%%%%%%%%%%%%%%%%%%%%
\section{Introduction}

Supernovae and hypernovae are among the most spectacular events in the
observable universe with an enormous amount of energy involved. Numerous
studies have been undertaken to understand their mechanisms, for recent
reviews, see e.g.~\cite{kota_06, ott_09,jank_12} and references therein. At
the origin of an iron core-collapse supernovae is the gravitational collapse
of the core of a massive progenitor star exceeding its Chandrasekhar mass. The
induced electron captures, reducing the degeneracy pressure of the core,
enable the collapse to proceed until the matter density is high enough for
nuclear forces to become repulsive. This is the case when, roughly, nuclear
matter saturation density has been reached. At this point, a bounce occurs,
leaving a compact remnant~: a proto-neutron star (PNS), or possibly a black
hole (BH).

The bounce creates a shock that propagates outwards and soon
stalls, having lost a lot of energy in photodissociation of iron group
nuclei within the infalling material. The shock is known to stall in
the semi-transparent regime, where the neutrinos are decoupled from
the fluid but can still interact. In particular, they deposit energy
behind the shock via charged-current interactions. While the
collapse, bounce and prompt shock propagation phases seem to be well
reproduced by spherically symmetric simulations, it is now widely
admitted that the late phases are deeply multidimensional.

Over decades much effort has been devoted to numerical simulations in order to
gain physical insight. Increasing computer power and refined models have lead,
in recent years, to considerable progress in understanding, in particular, the
supernova explosion mechanism using multidimensional models. Indeed, aided by
convection (e.g. \cite{murp_11}) and the standing accretion shock instability
(SASI) \cite{blon_03, fogl_06}, several authors reported explosions by the
neutrino heating mechanism in 2D or 3D (e.g. \cite{mare_09, suwa_10,
  brue_10,taki_12}). Although some drawbacks still exist, for instance the
reported explosion energies are low ($\sim 10^{50} \ \textrm{erg}$ compared
with the canonical observed value of $\sim 10^{51} \ \textrm{erg}$), reviving
the shock by depositing neutrino energy in the gain layer in a
multidimensional simulation seems to be a promising way to finally make the
supernovae explode (e.g. \cite{jank_12, kota_12b}). However, other mechanisms
like the acoustic mechanism (e.g. \cite{burr_06, jank_12}) or the QCD phase
transition mechanism \cite{sage_09} cannot be excluded. The MHD mechanism
(e.g. \cite{ober_09,ende_12,taki_11,wint_12}) could also account for the
explosion of rapidly rotating, highly magnetized cores and lead to luminous
supernovae or hypernovae.

Black hole formation has been intensely investigated (see
e.g.~\cite{sumi_07, fish_09, ocon_11,hemp_12}). A simulation in which the
shock is not able to recover positive velocities and to break through the
infalling material from the progenitor leads to the formation of a BH, which
often swallows the entire progenitor~\cite{ugli_12}. This is often referred to
as a failed supernova. Neutrino emission in failed supernovae simulations
stops abruptly when the emission region enters the apparent horizon, as one
could expect. This gives a good criterion to discriminate between BH formation
and neutron star formation in an upcoming neutrino signal~\cite{kota_12}. Due
to larger accretion rates, failed supernovae simulations are also known to
explore higher densities and temperatures than their exploding counterparts,
which makes them a tool to explore finite temperature equations of state
(EoSs) at supranuclear density.

The EoS remains one of the uncertainties in a stellar core-collapse numerical
simulation. The vast majority of simulations employs either the H.Shen
\textit{et al.}~EoS (HShen, \cite{shen_98}) or the Lattimer and Swesty EoS
(LS, \cite{latt_91}). Both of them assume the same particle content: free
nucleons, $\alpha$-particles, one (representative) heavy nucleus, electrons,
positrons and photons. Concerning the nuclear interaction, the HShen EoS is
based on a relativistic mean field model with a Thomas--Fermi approximation
for the description of the inhomogeneous part, while the LS EoS uses a
non-relativistic Skyrme-type model with a simplified momentum-independent
nucleon-nucleon interaction. The latter EoS is further described in
Sec.~\ref{section:eos}. EOS dependence has been studied since many years with,
in particular, studies of the effects of incompressibility in LS
EoS~\cite{thom_03}, or differences between HShen and LS for BH
formation~\cite{sumi_07, fish_09, ocon_11}. More recently, it has been
shown~\cite{suwa_12} that differences in the nuclear properties can account
for differences in the resulting hydrodynamics, specifically the predicted
central density, pressure or compactness of the proto-neutron star. Suwa
\textit{et al}~\cite{suwa_12} have also shown that there is a difference in
the gain layer where convection is triggered and the simulations behave
differently, qualitatively and quantitatively. These differences can be large
enough to govern the presence of an explosion.

Some other finite temperature EoSs are available, especially recent ones
(e.g. \cite{hemp_12, shen_11}), focusing on improvements in the subsaturation
regime. In this paper, we focus on a different aspect, not included in the
standard EoSs, namely the presence of additional particles at high densities
and temperatures. It is natural and widely accepted that additional particles
should appear in matter at densities above roughly nuclear matter saturation
density~\cite{latt_07}. The high temperatures during core-collapse are in favor of the
population of these additional states. Supernovae simulations with pions and
hyperons added to the HShen EoS were reported in Ishizuka \textit{et
  al.}~\cite{ishi_08}, Sumiyoshi \textit{et al}~\cite{sumi_09} and Nakazato
\textit{et al.}~\cite{naka_12}. The results show in particular that the time
from bounce to BH collapse is shortened by the presence of pions and/or
hyperons. The neutrino signal is hardly, or not at all changed, except for the
fact that the duration of the signal is shorter due to earlier collapse to a
BH. The reason is that pions and hyperons appear only deep inside the
PNS. Nakazato \textit{et al.}~\cite{naka_12} show in addition that a variation
of the coupling parameters of hyperons have only very little influence on
these qualitative results. We extend this work to the other commonly used EoS,
the LS EoS.

We discuss two extensions of the LS EoS: one including a free pion gas and
another one including the $\Lambda$-hyperon.  Unfortunately, very few
constraints on the interactions of these particles exist, be it from nuclear
physics experiments or astronomical observations. The most stringent one at
the moment is probably the observation of an almost $2 \ \textrm{M}_\odot$
neutron star~\cite{demo_10}, a mass measured with high precision by Shapiro
delay. Many EoSs including additional particles, in particular hyperons, are
in contradiction with this observation due to a too strong softening of the
EoS. For instance, with the HShen EoS + $\Lambda$~\cite{hshen_11} only a
maximum neutron star mass of $\sim 1.75 \ \textrm{M}_\odot$ is obtained. With
the Ishizuka {\it et al.}  EoSs \cite{ishi_08}, the maximum masses for their
EoSs with different parameterizations of hyperons, muons and pions are $1.55,
1.63$ and $1.65 \ \textrm{M}_\odot$. Our extensions of the LS EoS have been
described in detail in~\cite{oert_12}, where it was shown in particular that
it is possible, including hyperons in the EoS, to fulfill the constraint from
the $2 \ \textrm{M}_{\odot}$ neutron star measurement together with the
available experimental constraints on hyperon couplings. In
Sec.~\ref{section:eos} we shall introduce the EoSs models we are employing.

Our simple neutrino treatment does not allow to extract an
energy-resolved neutrino signal, we thus only discuss some results for
integrated luminosities and we will concentrate the discussion on the
time between bounce and black hole collapse as well as on the
properties of the different PNSs. The EoS with the $\Lambda$-hyperon
shows a phase transition~\cite{gulm_12b} to hyperonic matter, whose
imprints on the collapse shall be discussed.

This paper is organized as
follows. In Sec.~\ref{section:equations}, we review the equations solved by
our code and describe the employed finite temperature EoSs. In
Sec.~\ref{section:leakage} we describe the newly implemented leakage scheme,
and then our results are discussed in Sec.~\ref{section:results}.

In this paper we use a metric signature $(-, +, +, +)$ and geometrical
units in which $c = G = k_B = 1$. Greek indices run from 0 to 3, while Latin
indices run from 1 to 3. We adopt the Einstein summation convention, too.

%%%%%%%%%%%%%%%%%%%%%%%%%%%%%%%%%%%%%%%%%%%%%%%%%%%%%%%%

\section{Model and equations}
\label{section:equations}

We perform a series of numerical simulations using the \textsc{CoCoNuT} code
in spherical symmetry. Note that the code can be run in 2D/3D too, but due to
the very long simulation times, we restrict ourselves to 1D
simulations. Nevertheless, the model and equations presented here could be in
principle applied to 2D or 3D cases, without any loss of generality. Since
details are already available in the literature \cite{dimm_05, dimm_08,
  cord_09}, in this section we only briefly describe the equations solved by
the code. The newly implemented neutrino treatment is described in detail in
the next section.

\subsection{General relativistic hydrodynamics}
\label{subsection:grhydro}

General relativity (GR) is used through the 3+1 approach (see e.g.~\cite{3plus1}
and references therein), where the 4-metric $g_{\mu \nu}$ is described by the lapse function
$\alpha$, the shift 3-vector $\beta^i$ and the spatial 3-metric
$\gamma_{ij}$. The line element is then written as 

\begin{equation}
  \label{eq:metric}
  \mathrm{d}s^2 = -\alpha^2\mathrm{d}t^2 + \gamma_{ij} (\mathrm{d}x^i + \beta^i\mathrm{d}t) (\mathrm{d}x^j + \beta^j\mathrm{d}t)~.
\end{equation}

Here we use the isotropic gauge, in which the 3-metric is written as

\begin{equation}
  \label{eq:cfc}
  \gamma_{ij} = \Phi^4 f_{ij}~,
\end{equation}

with $f_{ij}$ denoting the flat-space 3-metric and $\Phi$ the
conformal factor. Although this gauge choice is fully valid in spherical
symmetry, it is denoted as \emph{conformally flat condition} (CFC,
see~\cite{wils_96}) in the general case (2D or 3D). This approximation misses some
aspects of GR (it does not contain gravitational waves, it cannot exactly
describe a Kerr black hole or a rotating fluid configuration), but it has been
proven to be a very good approximation in the case of core-collapse
simulations \cite{shib_04, cerd_05}. 

Assuming here a perfect fluid with 4-velocity $u^\mu$ and baryon
number density $n_b$, we can define a 4-current as $n_bu^\mu$. For
convenience, \textsc{CoCoNuT} works with the fluid density $\rho = m_n
n_b$, where $m_n$ is the neutron mass, so that the 4-current reads

\begin{equation}
  \label{eq:current}
  J^\mu = \rho u^\mu~.
\end{equation}

The stress-energy tensor is given by
\begin{equation}
  \label{eq:tmunu}
  T^{\mu \nu} = \rho h u^\mu u^\nu + P g^{\mu \nu},
\end{equation}
with $P$ being the fluid pressure and $h = (e + P)/ \rho$ 
the specific enthalpy; $e$ is here the energy density of the fluid.
These two quantities obey local conservation equations
\begin{equation}
  \label{eq:divj}
  \nabla_\mu J^\mu = 0\quad\mathrm{and}\quad  \nabla_\mu T^{\mu \nu} = 0~,
\end{equation}
%  \label{eq:divtmunu}
where $\nabla_\mu$ denotes the covariant derivative associated to the 4-metric
$g_{\mu \nu}$. The electron fraction $Y_e = n_e/n_b$, electron neutrino
fraction $Y_{\nu_e} = n_{\nu_e}/n_b$ and electron antineutrino fraction
$Y_{\bar{\nu}_e} = n_{\bar{\nu}_e}/n_b$ enter separately three advection
equations, which are included hereafter in a conservative-like
form.

No advection equation has been introduced for 
$Y_{\nu_x}$, with $\nu_x = \left\{ \nu_\mu, \nu_\tau, \bar{\nu}_\mu,
  \bar{\nu}_\tau \right\}$ standing for muon and tau neutrinos. Before
bounce, $Y_{\nu_x}$ vanishes to a very good 
approximation, because of the absence of charged-current production
processes due to the lack of heavy charged leptons. The only
production channels are pair productions (see
Sec.~\ref{sss:nu_creation}) which start to play a role in 
the hot environment close to and after bounce. After bounce, we
determine $Y_{\nu_x}$ from the integration of the Fermi-Dirac
distribution with zero chemical potential for the $\nu_x$, thus assuming
thermally equilibrated $\nu_x$. This simplification saves some
computational time and should not have any major influence on our
results, since non-electronic neutrinos are much less abundant than
electronic ones. Non-electronic neutrinos have thus only little effect
on the overall dynamics compared with the electronic ones (see
e.g.~\cite{lieb_09}, where the dynamics of a full Boltzmann simulation
is well reproduced by a simplified simulation with no $\nu_x$).

The 3+1 formalism slices the four-dimensional spacetime by three-dimensional
spacelike hypersurfaces, and it is therefore particularly well adapted to
describe hydrodynamics with respect to a Eulerian observer, which moves
orthogonally to these spacelike hypersurfaces with a 4-velocity $n^\mu$. 
The $3+1$ decomposition of the stress-energy tensor reads
\begin{equation}
  \label{eq:tmunu3p1}
  T^{\mu\nu} = E\,n^\mu n^\nu + n^\mu S^\nu + S^\mu n^\nu + S^{\mu\nu}, 
\end{equation}
$E$ being the matter energy density, $S^i$ the matter momentum density, and
$S^{ij}$ the matter stress tensor, all measured by the Eulerian observer. Note
that the momentum density and the matter stress tensors are orthogonal to the
Eulerian observer's 4-velocity $n^\mu$, they can be therefore described as
3-tensors (i.e. using Latin indices) tangent to the spacelike hypersurfaces. 

The Eulerian observer sees the fluid with a 3-velocity $v^i = u^i /
(\alpha u^0) + \beta^i / \alpha$. In order to simplify notations, we
also introduce the velocity $\hat{v}^i = v^i - \beta^i / \alpha$. To
write explicitely the hydrodynamic equations in flux-conservative
form~\cite{bany_97}, we define a set of ``conserved'' variables

\begin{equation}
  \label{eq:conserved}
  D = \rho W, \ \ S^i = \rho h W^2 v^i, \ \ \mathcal{E} = E - D = \rho h W^2 - P - D~,
\end{equation}
where $W = \alpha u^0 = 1 / \sqrt{1 - v_iv^i}$ is the Lorentz
factor. 

From Eqs.~(\ref{eq:divj}) and (\ref{eq:conserved}), adding the three
advection equations for $Y_e, Y_{\nu_e}, Y_{\bar\nu_e}$, we then derive
flux-conservative hyperbolic Eqs.~\cite{bany_97}:

\begin{equation}
  \label{eq:hydro}
  \frac{\partial \sqrt{\gamma}\bm{U}}{\partial t} + \frac{\partial
  \sqrt{-g}\bm{F}^i}{\partial x^i} = \sqrt{-g} \bm{\Sigma},
\end{equation}

where

\begin{eqnarray*}
  \bm{U} &=& [D, S_j, \mathcal{E}, DY_e, DY_{\nu_e},
  DY_{\bar{\nu}_e}]~, \label{eq:uhydro} \\
  \bm{F}^i &=& [D\hat{v}^i, S_j\hat{v}^i + \delta^i_jP,\mathcal{E}
  \hat{v}^i + Pv^i, DY_e\hat{v}^i, DY_{\nu_e}\hat{v}^i,
  DY_{\bar{\nu}_e}\hat{v}^i]~, \label{eq:fhydro} \\
  \label{eq:shydro}
  \bm{\Sigma} &= &[0, \frac{T^{\mu \nu}}{2} \frac{\partial g_{\mu
      \nu}}{\partial x^j},
     T^{00} \left( K_{ij}\beta^i\beta^j - \beta^i\frac{\partial
      \alpha}{\partial x^i} \right) +     \nonumber\\
    &&T^{0i} \left( 2K_{ij}\beta^j - \frac{\partial
      \alpha}{\partial x^i} \right) + 
  T^{ij}K_{ij},     \nonumber\\
    &&\Sigma_e,\Sigma_{\nu_e},\Sigma_{\bar{\nu}_e}]~.
\end{eqnarray*}

$g$ and $\gamma$ are the determinant of $g_{\mu \nu}$ and
$\gamma_{\mu \nu}$, respectively, and $K_{ij}$ is the extrinsic
curvature, defined by
Eq.~(\ref{eq:kij}). $\Sigma_e,\Sigma_{\nu_e},\Sigma_{\bar{\nu}_e}$ are
the creation-annihilation terms that shall be detailed hereafter. 
These general-relativistic hydrodynamic 
equations are solved using high-resolution shock-capturing schemes
(see e.g.~\cite{font-08} and references therein).

The creation-annihilation terms are computed as follow:
\begin{eqnarray}
  \label{eq:newse}
  \Sigma_e &=& - \Sigma_{ec} + \Sigma_{pc}~,\\
    \label{eq:newsve}
    \Sigma_{\nu_e} &=& \Sigma_{ec} + \Sigma_{ee} + \Sigma_{pl} - \frac{Y_{\nu_e}}{t_{esc}}~,\\
    \label{eq:newvbe}
  \Sigma_{\bar{\nu}_e} &=& \Sigma_{pc} + \Sigma_{ee} + \Sigma_{pl} - \frac{Y_{\bar{\nu}_e}}{t_{esc}}~,
\end{eqnarray}
where $\Sigma_{ec}$ denotes the electron capture rate, $\Sigma_{pc}$ the
positron capture rate, $\Sigma_{ee}$ and $\Sigma_{pl}$ the pair annihilation and
plasmon decay rates, respectively (defined in
Appendix~\ref{app:microphys}). We have introduced here the neutrino leak
term $Y_\nu / t_{esc}$ that represents the number of neutrinos able to
reach the neutrinosphere, i.e. able to escape. $t_{esc}$ is the time
needed for the neutrinos present at a given radius $r$ to reach the
neutrinosphere and is discussed in Sec.~\ref{sss:nu_creation}, too.

 Note the following modifications compared with
previous work~\cite{dimm_08}:
\begin{enumerate}
\item The neutrino pressure no longer appears in the hydrodynamic
  source terms, Eq.~(\ref{eq:hydro}). It is instead taken into
  account together with the other neutrino source terms, see
  Eq.~(\ref{eq:tmunusource}).
\item Two equations for $Y_{\nu_e}$ and $Y_{\bar{\nu}_e}$ are added to
  describe neutrino advection.
\item Three sources, $\Sigma_e,\Sigma_{\nu_e}$ and
  $\Sigma_{\bar{\nu}_e}$, are added. $\Sigma_{\nu_e}$ and
  $\Sigma_{\bar{\nu}_e}$ track creation and annihilation of neutrinos
  and, via lepton number conservation, $\Sigma_e$ accounts for the
  changes in the electron fraction.
\end{enumerate}

All these are further detailed in the Sec.~\ref{section:leakage} and
explicit expressions are given in App.~\ref{app:hydro_sources} and
App.~\ref{app:microphys}. Simulations are performed on a Eulerian grid
with spherical geometry. Our spherically symmetric runs have 805 ($r$)
points, and a logarithmic grid spacing.

Global convergence of the different parts of the CoCoNuT code have
been studied in details in \cite{dimm_05}. Our new numerical
developments, namely the small changes in the hydrodynamics presented
here and the leakage scheme, do not modify the convergence properties
for the gravitational and hydrodynamics solvers.

\subsection{Gravitational potential equations in isotropic gauge}

Our treatment of the gravitational potential is similar to
\cite{dimm_05, cord_09}. We briefly review here the set of equations
solved by \textsc{lorene}~\cite{lorene}, a C++ library designed to
solve 3+1 GR equations by means of spectral
methods~\cite{gran_09}. The \textsc{CoCoNuT} code uses this library
for solving gravitational field equations through the ``marriage des
maillages'' approach~\cite{dimm_05}.

We start by defining the extrinsic curvature

\begin{equation}
  \label{eq:kij}
  K^{ij} = \frac{1}{2\alpha \Phi^4}\left( \mathcal{D}^i \beta^j +
    \mathcal{D}^j \beta^i -
    \frac{2}{3}f^{ij}\mathcal{D}_k \beta^k \right),
\end{equation}

where $\mathcal{D}_j$ is the covariant derivative associated with the flat
metric $f_{ij}$. We can then introduce the quantity

\begin{equation}
  A^{ij} = \Phi^{10} K^{ij},
\end{equation}

and the momentum constraint can be written as

\begin{equation}
\label{eq:momconstr}
  \mathcal{D}_j A^{ij} = 8 \pi \Phi^{10} S^i = 8 \pi \Phi^{6} f^{ij}S_j = 8 \pi f^{ij}S_j^*,
\end{equation}

with $S_j^* = \Phi^{6}S_j$.

Now, we write $A^{ij}$ as
\begin{equation}
  \label{eq:aij}
  A^{ij} = \mathcal{D}^i X^j + \mathcal{D}^j X^i - \frac{2}{3}\mathcal{D}_k X^k f^{ij},
\end{equation}

where $X^i$ is a 3-vector (see \cite{cord_09}, too). Combining
Eqs.~(\ref{eq:momconstr}) and (\ref{eq:aij}) we
get

\begin{equation}
  \label{eq:grav1}
  \Delta X^i + \frac{1}{3}\mathcal{D}^i\mathcal{D}_jX^j = 8 \pi f^{ij}S_j^*,
\end{equation}

which is the first equation solved by the code to obtain $X^i$; we can then
deduce $A^{ij}$ thanks to Eq.~(\ref{eq:aij}) and solve the
Hamiltonian constraint equation

\begin{equation}
  \label{eq:hamilconstr}
  \Delta \Phi = -2\pi \Phi^{-1}E^* - \Phi^{-7} \frac{f_{il}f_{jm}A^{lm}A^{ij}}{8},
\end{equation}

where $E^* = \Phi^6 E$. Knowing the conformal factor, we
can solve

\begin{equation}
  \label{eq:phialpha}
    \Delta (\Phi \alpha) = 2 \pi \alpha \Phi^{-1}(E^* + 2S^*) + \alpha \Phi^{-7}
  \frac{7f_{il}f_{jm}A^{lm}A^{ij}}{8},
\end{equation}

with $S^* = \Phi^6 \gamma_{ij} S^{ij} = \Phi^6 \left( \rho h (W^2 - 1) +
  3P\right)$ the conformally rescaled trace of the matter stress tensor. It is
thus possible to get the lapse. Then, by solving 

\begin{equation}
  \Delta \beta^i + \frac{1}{3}\mathcal{D}^i(\mathcal{D}_j\beta^j) = \mathcal{D}_j(2 \alpha \Phi^{-6} A^{ij}),
\end{equation}

we obtain the shift. Note that this approach is not bound to the
spherically-symmetric case (see~\cite{cord_09} in the 3D and CFC case).

\subsection{Apparent horizon}

We use the apparent horizon (AH) finder described in details in
\cite{lin_07}. In a simulation, a PNS accretes mass and looses energy due to
neutrinos at the same time. This enables the PNS to further contract, and
consequently the density increases. At some point, depending on the EoS, the
PNS becomes gravitationally unstable and collapses to a BH. The BH collapse is
unambiguous, as density, temperature, pressure, internal energy all increase
very rapidly. Since the lapse is decreasing from $\alpha \simeq 1 $ in the
Newtonian limit to zero at the center, some authors just take a given value
$\alpha \ll 1$ of the lapse as a criterion for BH formation (see
e.g.~\cite{fish_09}). However, using an AH finder that detects a marginally
trapped surface is a more rigorous way to treat BH formation. In addition, the
AH finder enters a simulation only at the very end, when BH collapse begun,
triggered by a chosen value for the conformal factor at the center (typically,
$\Phi(r=0) \geq 2$), it is thus computationally cheap.

\subsection{Initial models}
\label{section:initialmodels}

We employ two different 40 $\textrm{M}_\odot$ (zero-age main sequence (ZAMS)
mass) progenitors. As a first progenitor model, we have chosen a star with low
metalicity ($10^{-4}$ solar) from Woosley, Heger, and Weaver~\cite{woos_02}
(hereafter u40). The low metalicity leads to a much higher accretion rate
after bounce compared with a solar metalicity star. Therefore, even though
there still are many unknowns concerning the fate of a particular progenitor
(see e.g.~\cite{ugli_12}), very massive stars with low metalicity are widely
accepted as good candidates for BH formation.

The second one, from Woosley and Weaver~\cite{woos_95} (hereafter WWs40),
has solar metalicity. This progenitor was widely used in BH collapse
studies (see e.g.~\cite{ocon_11, fish_09, sumi_07, hemp_12}) and we
use it here mainly for comparison purposes. It is believed to collapse
to a BH, as the accretion rate is too high and the iron core is too
massive for the star to explode. Let us stress, however, that Ugliano
{\it et al}~\cite{ugli_12} report that the $40 \ \textrm{M}_\odot$
progenitor with solar metalicity from Woosley, Heger, and
Weaver~\cite{woos_02}, does explode and leaves a NS rather than a
BH. The reason is probably that it has a smaller accretion rate than
its counterpart from~\cite{woos_95}, a difference that may come from
the different treatment of the mass loss.

%%%%%%%%%%%%%%%%%%%%%%%%%%%%%%%%%%%%%%%%%%%%%%%%%%%%%%%%

\subsection{Equations of state}
\label{section:eos}
In the central region, i.e. mainly within the hot PNS, very high
densities (roughly above nuclear saturation) and temperatures (several
tens of MeV) are reached. Under these conditions, the particle content
of the standard nuclear EoS such as the HShen EoS or the LS EoS is
possibly not sufficient. As already mentioned, the latter models the
matter as a mixture of one (average) heavy nucleus,
$\alpha$-particles, free nucleons, electrons, positrons and photons.
Additional particles, such as thermal pions, hyperons, or nuclear
resonances should appear. Even a QCD phase transition is
conceivable~\cite{sage_09}. In the last years some work in this
direction has begun, extending the HShen EoS to include pions or
hyperons, see e.g.~\cite{sumi_09, ishi_08,hshen_11}, or quarks,
see~\cite{sage_09}.

\begin{table*}
  \caption{Properties of cold neutron stars for the different employed
    EoSs. The LS180 EoS is added for comparison. The given radii are
    circumferential ones.  
    \label{table:propcoldns}
  }
  \begin{center}
    \begin{tabular}{ccccc}
      \hline \hline
      Name & LS180 & LS220 & LS220+$\Lambda$ & LS220 + $\pi$ \\
      \hline
      Maximum gravitational mass [$M_\odot$] &  1.84    & 2.06 & 1.91 & 1.95\\
      Maximum baryonic mass [$M_\odot$] &  2.12    & 2.40 & 2.22 & 2.27\\
      
      Radius at maximum mass [km] & 10.13 & 10.67& 9.28
      & 9.94 \\
     
      Radius at $M_g = 1.4 M_\odot$ [km] &  12.19   & 12.71      & 12.41    
      &  12.03 \\
      Central density at maximum mass [fm$^{-3}$]&  1.26    & 1.09      & 1.47 
      &  1.26   \\
      \hline \hline
    \end{tabular}
  \end{center}
\end{table*}

Here we employ EoSs which are extensions of the other widely used nuclear
EoS by Lattimer and Swesty~\cite{latt_91}. Electrons and positrons are
treated as non-interacting relativistic gas in pair equilibrium,
neglecting electron-screening effects; photons are treated as an ideal
ultra-relativistic gas. Equilibrium with respect to strong and
electromagnetic interactions is assumed, while $\beta$ equilibrium is
not requested, which is consistent with expectations during
core-collapse events. Nuclear interaction is treated using a
liquid drop model and the transition between inhomogeneous and
homogeneous nuclear matter is described via a Maxwell
construction. Details can be found in the original
paper~\cite{latt_91}.

There are three different sets of parameter values available,
resulting in three different values of the nuclear incompressibility,
$K = 180, 220,$ and $375$ MeV. In this work, we restrict ourselves to
the value $K = 220$ MeV, which gives an EoS compatible with both
nuclear data, which suggest a value around $K = 240 \textrm{MeV}$ (see
e.g.~\cite{shlo_06}), and the recent high mass neutron star of $M =
1.97 \pm 0.04 \ \textrm{M}_\odot$~\cite{demo_10}. The maximum mass of
a cold neutron star is 2.06 $\textrm{M}_\odot$ (2.40
$\textrm{M}_\odot$ baryonic mass) with $K = 220$ MeV. The EoS with $K = 180$
MeV, on the lower end of possible $K$-values from nuclear data, fails
to reproduce the neutron star mass constraint giving a maximum cold
neutron star mass of only $1.84 \ \textrm{M}_\odot$ and $K = 375$ MeV
lies well above the allowed region from nuclear data. For a thorough
comparison of the three different LS EoSs in the context of core
collapse simulations, see~\cite{suwa_12}.
 
In~\cite{oert_12} different possible extensions of the LS EoS are
discussed, considering hyperonic degrees of freedom as well as pions
and muons. Within this work we limit the discussion to two
rather simple cases: one including pions and the other one including
the $\Lambda$-hyperon. For the former one, pions have been added upon
the LS EoS as a free gas. For a critical discussion of this
approximation see~\cite{oert_12}.

The second one, including the $\Lambda$-hyperon can be seen as
analogue to the HShen + $\Lambda$ EoS from~\cite{hshen_11}. The
motivation is that the $\Lambda$ represents, together with the
$\Sigma^-$-hyperon, probably the most important hyperonic degree of
freedom in hot dense supernova matter. Thus, including the $\Lambda$
allows for discussing general features of the effects coming from the
hyperonic degrees of freedom, without the necessity of resolving the
complicated particle composition in the presence of many different
hyperons. Indeed, most of them have very low abundances, see
e.g.~\cite{oert_12}. The $\Lambda\Lambda$ and the $\Lambda N$
interactions are taken from the model by Balberg and
Gal~\cite{balb_97} with the parameterization 220BG
from~\cite{oert_12}. This model has the advantage of matching well
with the LS EoS: in the region where no $\Lambda$-hyperons are
present, the interaction is exactly the same as in the LS model.

Within the Balberg and Gal model, the
$\Lambda$-hyperons appear through a first order phase transition,
see~\cite{gulm_12a, gulm_12b}. A Gibbs construction is employed to
describe matter in the phase coexistence region, see~\cite{gulm_12b}
for details. The effect of the criticality of the phase transition on
the simulation shall be the subject of forthcoming work.

The maximum gravitational masses of cold neutron stars are $M = 1.95 \
\textrm{M}_\odot$ for the LS220+$\pi$ EoS ($2.27 \ \textrm{M}_\odot$ baryonic
mass) and $M= 1.91 \ \textrm{M}_\odot$ for the LS220+$\Lambda$ EoS ($2.22 \
\textrm{M}_\odot$ baryonic mass) with parameterization 220BG. The former value
lies in the 1$\sigma$-range for the mass constraint from PSR J 1614-2230, the
latter is only very slightly below. Properties for cold neutron stars for the
different EoSs used within this work are summarized in
Table.~\ref{table:propcoldns}. The LS model with $K = 180$ MeV is added for
comparison. The circumferential radii at the canonical value of $M = 1.4 \
\textrm{M}_\odot$ vary only very little for the four EoSs, whereas the radii
at maximum mass are lower for the EoSs with additional particles, as expected
from the softer character of these EoSs.

%%%%%%%%%%%%%%%%%%%%%%%%%%%%%%%%%%%%%%%%%%%%%%%%%%%%%%%%

\section{Leakage scheme}
\label{section:leakage}

\subsection{Introduction}

Neutrino transport is one of the most challenging aspects of modern
supernovae simulations. Because the shock stalls in the
semi-transparent regime, an accurate neutrino treatment should rely,
in principle, on solving the Boltzmann equation. Several authors
report on supernovae simulations with neutrino transport, either
Newtonian (e.g. \cite{bura_06, ott_08}), or general relativistic
(e.g. \cite{lieb_04, muel_10}), but this remains computationally
challenging~: a simulation with 3D GR hydrodynamics and a (6D)
Boltzmann solver is not done yet (see also \cite{kota_12}). The use of
simplified models is therefore fairly common. They are used to tackle
3D hydrodynamics simulations \cite{nord_10}, or for parametric
studies of black hole collapse, which is known not to require a fully
detailed neutrino transport \cite{ocon_11, ugli_12}. In this
section, we provide details concerning the implementation of our
leakage scheme, mostly inspired by \cite{ruff_96, ross_03, ocon_10,
  seki_10}.

The main idea of the leakage scheme is to treat the neutrinos as a fluid
component inside the neutrinosphere. Above the latter, neutrinos are
considered as free streaming, and only the energy taken away from the fluid
needs to be considered. This is achieved by adding terms to the sources of
hydrodynamic equations, Eqs.~(\ref{eq:hydro}). These additional source terms
are explicitly given in Appendix~\ref{app:hydro_sources}. It is a grey scheme,
that defines and works with a mean energy for every neutrino species. The
leakage scheme is, by construction, not sufficient to treat the
semi-transparent regime nor to revive the shock. On the other hand, it is a
simplified model that comes from physical arguments and enables us to run
core-collapse simulations within a reasonable amount of
time\footnote{e.g. fiducial model lsu40 takes 152 minutes for a
  single-processor run. The total physical time simulated in the code is
  871ms.}.

\subsection{Implementation}

\subsubsection{Neutrinosphere}
\label{section:nusph}

The neutrinosphere is the limit where neutrinos decouple from the fluid. This
is characterized by an optical depth $\tau$ of the order of unity. We choose
to define it as the region where $\tau = 2/3$, consistently with the
literature (see e.g. \cite{brue_85, ocon_10}). In order to find the exact
position of the neutrinosphere during a simulation, at each time step the
opacity is evaluated, using the mean energy of the neutrinos,
Eq.~(\ref{eq:epsnu}).

Following~\cite{ruff_96, ross_03, ocon_10, burr_06}, the neutrino reactions we
consider for calculating the opacity are the following (see also
App.~\ref{app:microphys} for the details of the implementation) :

\begin{enumerate}

\item Elastic scattering off a nucleon,
\begin{equation}
  \label{eq:nscatt}
  \nu_i + N \rightarrow \nu_i + N~,
\end{equation}

where $N = n, p$, and $\nu_i$ represents one of the three neutrino species
implemented : electron neutrino $\nu_e$, electron antineutrino $\bar{\nu}_e$
or other species $\nu_x$. 

\item Elastic scattering off a nucleus, 
\begin{equation}
  \label{eq:ascatt}
  \nu_i + (A, Z) \rightarrow \nu_i + (A, Z),  
\end{equation}
where $(A, Z)$ is the mean nucleus. 

\item Absorption of a $\nu_e$ by a neutron
\begin{equation}
  \label{eq:nabs}
  \nu_e + n \rightarrow p + e^-
\end{equation}
and absorption of a $\bar{\nu}_e$ by a proton
\begin{equation}
  \label{eq:pabs}
  \bar{\nu}_e + p \rightarrow n + e^+,
\end{equation}

are taken into account, when no $\beta$-equilibrium is assumed.

\item Elastic scattering off a $\Lambda$ 
  \begin{equation}
    \nu_i + \Lambda \rightarrow \nu_i + \Lambda~.
  \end{equation}
  We implement the coherent scattering off $\Lambda$-hyperons and take
  it into account in the opacity calculation, because the hyperon
  fraction can become quite large at the end of a simulation (see
  Sec.~\ref{section:results} for the hyperon fraction and
  App.~\ref{app:microphys} for the implementation). We study the
  impact of this newly implemented reaction on our simulations in
  Sec.~\ref{section:inflscatt}

\end{enumerate}
The optical depth is obtained from the total opacity, $\kappa$,  by integrating over
$r$,
\begin{equation}
  \label{eq:optdepth}
  \tau = \int_r^\infty \kappa \ \mathrm{d}r~.
\end{equation}
Note that this expression implicitly assumes that neutrinos only move along
radial rays.  In our code, the upper bound of the integral is taken to be the
last cell, where we start the integration, then we move inwards. The first
cell reaching $\tau \ge 2/3$ is taken to be the neutrinosphere.

As a refinement of this basic leakage scheme, an effective neutrino
chemical potential $\mu_{\nu,\mathit{eff}}$ is introduced in the
calculation of the different cross sections following~\cite{ruff_96}, 
\begin{equation}
  \label{eq:munuiter}
  \mu_{\nu,\mathit{eff}} = \mu_{\nu, eq} (1 - \exp(-\tau))~,
\end{equation}
with $\mu_{\nu, eq}$ being the neutrino chemical potential in
$\beta$-equilibrium given by the EoS as $\mu_{\nu, eq} = \mu_e + \mu_p
- \mu_n$. For matter completely transparent to neutrinos with a
vanishing optical depth, the effective chemical potential goes to
zero. In the other extreme case of a very large optical depth,
corresponding to matter opaque to neutrinos, the equilibrium value is
reached. Thus, it allows partly to correct for deviations from
equilibrium in the semi-transparent regime, although the neutrino
distribution function is still represented by a Fermi-Dirac
distribution.

From a practical point of view, Eq.~(\ref{eq:munuiter}) usually needs
very few iterations to converge (no more than 3), given the neutrino
effective chemical potential from the previous time step as a first
guess. At the beginning of a simulation, $\mu_{\nu,\mathit{eff}}$ is taken to
vanish everywhere.

\subsubsection{Neutrino creation and advection}
\label{sss:nu_creation}
Neutrinos are created by electron capture on free protons
(see App.~\ref{app:microphys} for the details of the implementation)
\begin{equation}
  \label{eq:eleccapt}
  p + e^- \rightarrow \nu_e + n,
\end{equation}

electron capture on nuclei
\begin{equation}
  \label{eq:eleccaptA}
  (A, Z) + e^- \rightarrow (A, Z-1) + \nu_e ,
\end{equation}

positron capture on free neutrons
\begin{equation}
  \label{eq:pcapt}
  n + e^+ \rightarrow \bar{\nu}_e + p,
\end{equation}

electron-positron pair annihilation
\begin{equation}
  \label{eq:pairann}
  e^- + e^+ \rightarrow \nu_i + \bar{\nu}_i,
\end{equation}

and plasmon decay
\begin{equation}
  \label{eq:plasm}
  \tilde{\gamma} \rightarrow \nu_i + \bar{\nu}_i.
\end{equation}
Since they appear only in the very dense regions close to the center,
where $\beta$-equilibrium is assumed, no charged-current reactions on
$\Lambda$-hyperons are considered.
 
%Electron-positron pair annihilation and plasmon decay are computed
%following~\cite{ruff_96}. The rates for electron and positron captures
%are computed following~\cite{brue_85}, integrated over neutrino
%energy. Therefore, they can be tabulated as a function of $\rho, T,
%Y_e$ (at each point of the EoS), and the effective neutrino chemical
%potential $\mu_{\nu,\mathit{eff}}$. Access to the tables is done by a
%quadrilinear interpolation.

Let us stress several approximations applied in the treatment of electron
and positron captures. We can justify these approximations by the fact
that the leakage by itself is a rather crude approximation, and going
into further details without a better neutrino treatment may not be
relevant or possible.  

Following the rates given in Bruenn~\cite{brue_85}, electron capture
on nuclei is cut off due to the assumption of shell closure for a
neutron number $N \ge 40$. It has been shown~\cite{lang_03} that
temperature effects smear out the shell structure and that the capture
rate on nuclei with $N \ge 40$ is therefore nonzero.

Electron and positron capture are only relevant out of
$\beta$-equilibrium, since in equilibrium the creation and absorption
terms cancel each other. Within the leakage scheme, however, the
equilibration cannot be described as the balance between both would
require the knowledge of the distribution function. We therefore take
only the creation part into account below a given critical density
where $\beta$-equilibrium is assumed. We choose this density to be
$1.17\times 10^{12} \mathrm{g}/\mathrm{cm}^3$ (see
Sec.~\ref{section:param}), which is in agreement with the general
finding that $\beta$-equilibrium sets in at a density of the order
$\rho \sim 10^{12} \mathrm{g}/\mathrm{cm}^3$, see
Sec.~\ref{section:thequil}, too.

In the region behind the prompt shock, matter is strongly neutronized,
giving very low values of $Y_e$. This effect is in general
overestimated by leakage schemes (see Fig.~\ref{comp_y_e_neutronis},
or e.g.~\cite{ocon_10}), among others due to the fact that the
absorption term is absent, which is slowing down electron capture in
this region close to $\beta$-equilibrium. In order to avoid the
leakage scheme giving (unrealistic) values of $Y_e$ below the lowest
one available in the EoS, electron capture is blocked below $Y_e =
0.045$. This allows to follow a simulation during and after the prompt
shock propagation

The reaction rates for neutrino creation and absorption enter the
sources of the new advection equations related to electron and
neutrino number in Eqs.~(\ref{eq:newse})-(\ref{eq:newvbe}). They are
detailed in App.~\ref{app:microphys}. We give here the
expression for the neutrino escape time $t_{esc}$, defined as
in~\cite{ruff_96}: 
\begin{equation}
  \label{eq:tesc}
  t_{esc} = a_1 (R_\nu - r) \tau~.
\end{equation}
$R_\nu$ is here the radius of the neutrinosphere. Ruffert \textit{et
  al.}~\cite{ruff_96} have set the free parameter $a_1$ to $a_1 = 3$,
in order to reproduce well data from transport calculations. We keep
this value in our implementation (see Sec.~\ref{section:param}).
 
Since we determine the position of the neutrinosphere at each time step, some
neutrinos may be no longer trapped at a given time step.  We separately
account for these freed neutrinos due to the inward movement of the
neutrinosphere. The treatment of energy losses due to escaping neutrinos will
be presented in Sec.~\ref{sec:energy}.

Finally, note that Eq.~(\ref{eq:tesc}) as well as
Eq.~(\ref{eq:optdepth}) do not have GR corrections. Neutrinos follow
a straight line (on the grid) while in principle they should follow a null
geodesic. This introduces a small numerical error, compared
with the overall approximations induced by the usage of a leakage scheme. In
particular, the \textit{ad hoc} factor of $a_1 = 3$ in Eq.~(\ref{eq:tesc}) is
already larger than the GR corrections in
Eqs.~(\ref{eq:tesc},\ref{eq:optdepth}) during our simulations. The biggest
underestimation of opacities occurs during BH formation, because the
neutrinospheres move closer to the center (or to the newly formed apparent
horizon). But we note that we are able to reproduce the collapse of the
PNS, which is essentially a free fall. 

\subsubsection{Deviation from thermal equilibrium}
\label{section:thequil}

During a core collapse event, neutrinos are in general not in thermal
equilibrium, which results in a distribution function different from
the Fermi-Dirac one, when using a realistic neutrino transport
scheme. Only in some very small regions it can be approximated by the
latter. The simplicity of the leakage does not allow for determining
the correct distribution function, implying that the neutrino number
given by integrating the Fermi-Dirac distribution function is not
the same as that given by advection and creation-absorption. Within
some leakage schemes the neutrino number is chosen to be given by
integration of the Fermi-Dirac distribution function, assuming thus
thermal equilibrium at all times. We have chosen to compute instead
the neutrino fractions from source terms and advection
equations. Therefore, the current neutrino number is not necessarily
the same as the value at thermal equilibrium, which is taken only as a
maximal value, similarly to~\cite{seki_10}.

\begin{figure}
 \centering
  \includegraphics[scale=0.7]{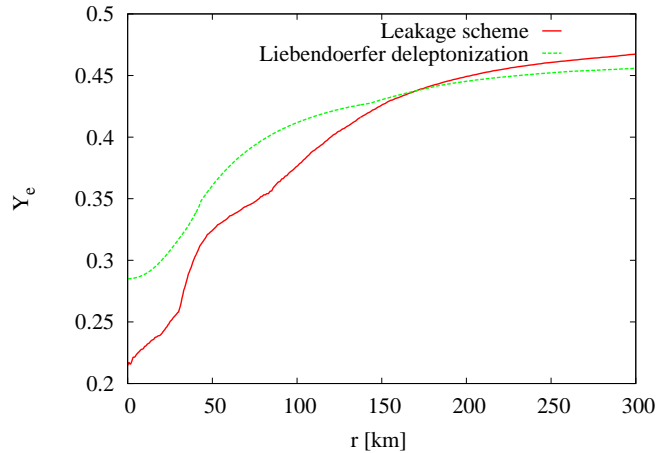}
  \caption{$Y_e$ as a function of radius at bounce, using the leakage
    scheme (plain, red) compared to the Liebend\"orfer deleptonization
    prescription (dashed, green)}
\label{comp_y_e_bounce}
\end{figure}

\begin{figure}
 \centering
  \includegraphics[scale=0.7]{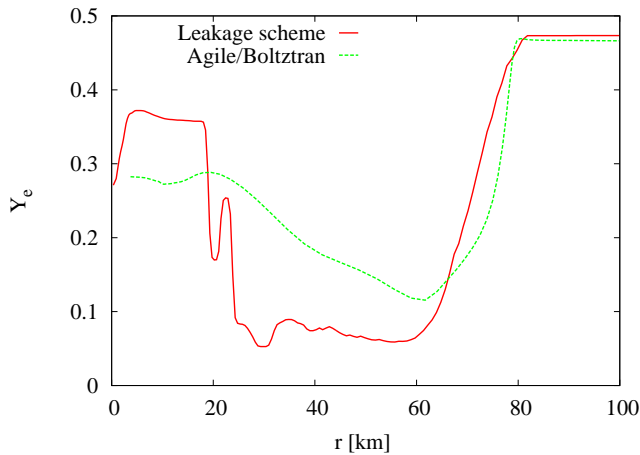}
  \caption{$Y_e$-profile during neutronization at $t=10 \ \textrm{ms}$ post-bounce
    using the leakage scheme (plain, red) compared with \textsc{Agile/Boltztran}
    (dashed, green).}
\label{comp_y_e_neutronis}
\end{figure}

This choice allows to perform simulations through the collapse and
bounce phases, where otherwise the neutrino number would be strongly
overestimated. To illustrate this, in Fig.~\ref{comp_y_e_bounce} we
display the $Y_e$-profile at bounce, a few milliseconds before the
launch of the shock. The result obtained with our leakage scheme is
compared with that employing the so-called Liebend\"orfer
deleptonization (see~\cite{lieb_05}). The latter is a parameterization
of full Boltzmann simulations and reproduces thus well a realistic
profile. We can see that the result of the leakage is in good
agreement with the predicted deleptonization. The deviation is at most
around 5\%, excepted at the very center where electron captures in the
Liebend\"orfer scheme are slowed down and finally stopped due to the
approaching and onset of $\beta$-equilibrium. $Y_e$ becomes constant
at a value around $0.28$. The leakage scheme, which neglects the
absorption term for electron capture cannot reproduce this
trend. $Y_e$ continues to decrease until the critical density for the
onset of $\beta$-equilibrium is reached. Since this density value is
adjusted to reproduce the $Y_e$-profile as well as possible during
neutronization after bounce (see Fig.~\ref{comp_y_e_neutronis} and
discussion below), $Y_e$ at the center at bounce is a little low ($Y_e
\approx 0.22$), but this should not have a significant impact on the
overall dynamics (see~\cite{ocon_12}, too).

Fig.~\ref{comp_y_e_neutronis} shows a comparison of the $Y_e$-profile
in the leakage scheme with that from the code \textsc{Agile/Boltztran}
(from Liebend\"orfer \textit{et al.}~\cite{lieb_05b}) with a Boltzmann
neutrino solver at 10 ms after bounce. This corresponds to the period
where matter behind the propagating prompt shock is strongly
neutronized. Note that, in order to have the same setting
as~\cite{lieb_05b}, we show here a spherically symmetric simulation
with a $15 \ \textrm{M}_\odot$ progenitor from Woosley and
Weaver~\cite{woos_95} with solar metalicity. Moreover, in contrast to
all other runs we have performed, the EoS used here is the LS EoS with
an incompressibility of $K = 180$ MeV, to match the setup of the run
with \textsc{Agile/Boltztran} from~\cite{lieb_05b}.

We observe that, as mentioned above, one of the drawbacks of the
leakage scheme is to overestimate the neutronization. The main reason
is that the equilibration of electron capture is not correctly
described. Otherwise, the leakage scheme is in qualitative agreement
with \textsc{Agile/Boltztran}. Note that the irregularities in $Y_e$
are due to the different simplifications of the leakage scheme.

\subsection{Treatment of energy losses}
\label{sec:energy}
In this section, we describe how the energy lost by the fluid, because
of neutrino emission, is taken into account. No heating is
implemented, as the leakage is not a transport scheme, and therefore
no self-consistent derivation of heating sources exist. Moreover,
heating does not play a major role in our simulations, given that we
expect a collapse to a BH (see discussion in
Sec.~\ref{section:initialmodels}).

Neutrinos enter the fluid momentum and energy equations as a source
term in the equation $\nabla_\beta T^{\alpha \beta} = 0$,

\begin{equation}
  \label{eq:tmunusource}
  \nabla_\beta T^{\alpha \beta} = q^\alpha,
\end{equation}

where we define the source 4-vector $q^\alpha$ as 

\begin{equation}
  \label{eq:qalpha}
  q^\alpha_{LF} = \left( \begin{array}{c} Q_E \\ Q_{(M)}^r \\
      Q_{(M)}^\theta \\ Q_{(M)}^\varphi  \end{array} \right).
\end{equation}

If the full neutrino distribution function were known as the solution of the
Boltzmann equation, the latter could be employed to build a consistent
neutrino stress-energy tensor from which the hydrodynamic sources $q^\alpha$
could be derived relying on the condition that the total stress-energy tensor
is divergence-free. This has been done in GR by several authors (see
e.g.~\cite{lieb_04, shib_11, muel_10}), incorporating the energy losses in a
coherent way. This is, however, not possible in a leakage scheme, where the energy
losses can only be implemented in an approximate way. We follow the choice
of~\cite{ocon_10} for the vector $q^\alpha$, which is detailed below. The
quantities in Eq.~(\ref{eq:qalpha}) are defined in the fluid rest frame (or
Lagrangian frame, hence the subscript LF), we further need to transform them
to the Eulerian frame where the hydrodynamic equations, Eqs.~(\ref{eq:hydro}),
are solved. This transformations are detailed in
Appendix~\ref{app:hydro_sources}, where the explicit expressions for the
source terms used in the code are given as well.

\subsubsection{Free streaming regime}

In the free streaming regime all neutrinos that are produced leave the
simulation, and take away their energy from the fluid. Knowing the neutrino
creation rates, $R_i$, from each considered reaction, and the mean neutrino
energy in each reaction, $\langle\epsilon_i\rangle$, (following
again~\cite{ruff_96}), we can write the total rate of energy loss by the fluid
in the free streaming regime as follows:

\begin{equation}
  \label{eq:qfreestr}
  Q_E = \Sigma_{ec} \langle \epsilon_{ec} \rangle + \Sigma_{pc} \langle
  \epsilon_{pc} \rangle + \Sigma_{ee} \langle \epsilon_{ee} \rangle +
  \Sigma_{pl} \langle \epsilon_{pl} \rangle + Q_{freed}~.
\end{equation} 

We have added here to the energy loss by neutrino creation in
electron/positron capture, pair annihilation and plasmon decay, the energy
loss rate, $Q_{freed}$, due to the freed neutrinos (if any) when the
neutrinosphere moves inwards. This term is new and comes from our
implementation of advection equations for neutrino fractions. See
App.~\ref{app:microphys} for further details on the implementation.  Note that
no momentum is removed from the fluid in the free streaming regime.

\subsubsection{Trapped regime}

In the trapped regime only those neutrinos reaching the neutrinosphere take
away energy. The total energy per unit time lost by the fluid in the trapped
regime thus reads

\begin{equation}
  \label{eq:qtrap}
  Q_E = - a_2 \langle\epsilon_\nu\rangle \frac{Y_\nu}{t_{esc}}~,
\end{equation}
with the mean energy per particle of escaping neutrinos, $\langle
\epsilon_\nu\rangle$. 
In our grey scheme, this mean energy per particle can be evaluated as follows,

\begin{equation}
  \label{eq:epsnu}
  \langle\epsilon_\nu\rangle  =   \frac{\int_0^\infty E f_{FD}(\eta)
    \mathrm{d}^3p}{\int_0^\infty f_{FD}(\eta) \mathrm{d}^3p} = T \frac{F_3 (\eta)}{F_2 (\eta)},
\end{equation}

where $f_{FD}$ is the Fermi-Dirac distribution function and $F_k$ is the Fermi
integral of order $k$, 

\begin{equation}
  \label{eq:fermi_integral}
  F_k (\eta) = \int_0^\infty \mathrm{d} x \ \frac{x^k}{1 +
\exp(x - \eta)}~.
\end{equation}

$T$ is the temperature and $\eta$ is the degeneracy parameter
($\eta_\nu = \mu_\nu / T$). Note that since
$\langle\epsilon_\nu\rangle$ depends on $\eta_\nu$, its value is
sensitive to the use of the effective chemical potential introduced in
Eq.~(\ref{eq:munuiter}).

The energy loss rate, Eq.~(\ref{eq:qtrap}), is calculated separately
for each neutrino species considered ($\nu_e$, $\bar{\nu}_e$,
$\nu_x$), each one having its own mean energy and its own fraction.

In order to correct partly for the fact that the leakage scheme
describes the energy loss only in an approximate way, we choose to
vary $a_2$ as a free parameter of order unity. It is adjusted to have
a mass of the PNS at BH formation comparable with the literature
\cite{fish_09, ocon_11}. In this paper, we fix $a_2$ to the value
$a_2 = 1.1$, see Sec.~\ref{section:param}.

In the trapped part, the momentum removed from the fluid is
approximated as the gradient of the neutrino pressure, $P_\nu$
~\cite{muel_10, ocon_10}. To compute it, neutrinos are assumed to
be an ideal Fermi gas of massless particles up to the
neutrinosphere. The neutrino pressure then reads

\begin{equation}
  \label{eq:press}
  P_\nu = \frac{4 \pi T^4}{3(hc)^3} (F_3(\eta_\nu) + F_3(-\eta_\nu)).
\end{equation}

The term proportional to $F_3(\eta_\nu)$ in Eq.~(\ref{eq:press}) takes
into account the $\nu_e$ pressure, while the one proportional to
$F_3(-\eta_\nu)$ takes into account the $\bar{\nu}_e$ pressure. The
pressure of $\nu_x$ is negligible and therefore not considered.

The momentum taken away from the fluid is then obtained by computing the
gradient of $P_\nu$ via~\cite{dimm_08, ocon_10}:

\begin{equation}
  \label{eq:qm}
  Q_{(M)i} = - \frac{\partial P_\nu}{\partial x^i};
\end{equation}

This expression shall be used to derive the neutrino energy-momentum source
terms for the hydrodynamic equations in Appendix~\ref{app:hydro_sources}.

Note that the components $Q_{(M)i}$ are exactly zero in the free streaming
regime.

\subsection{Summary of parameter values}
\label{section:param}
Our leakage scheme has three different parameters.
\begin{itemize}
\item The $\beta-$equilibrium density is taken to be $1.17 \times
  10^{12} \ \textrm{g.cm}^{-3}$ throughout this paper. Different
  authors report $\beta-$equilibrium to set in between $6 \times
  10^{11}$ and $2 \times 10^{12} \ \textrm{g.cm}^{-3}$~\cite{lieb_05,jank_07}. Our value is in
  agreement with this range. 
\item $a_1$, defined by Eq.~(\ref{eq:tesc}), has the numerical
  value $a_1 = 3$. We keep the value reported in~\cite{ruff_96}.
\item $a_2$, defined by Eq.~(\ref{eq:qtrap}), has the numerical
  value $a_2 = 1.1$. Since this parameter corrects for the
  incompleteness of the leakage scheme, it has to be close to
  unity. 
It is adjusted to have a mass of the PNS at BH formation
  comparable with the literature \cite{fish_09, ocon_11}.
\end{itemize}

The quantitative values found in our simulations are quite sensitive
to these parameters. Namely, a difference of $\pm 10\%$ in the
parameter $a_2$ leads to a difference in the PNS maximum mass of $\sim
2\%$, which can lead to significantly more time spent in the accreting
phase. A difference of $\pm 10\%$ in the $\beta-$equilibrium density
does not change the AH detection time, but the maximum density at
bounce changes by about $\pm 10\%$. These tests were done with the lsu40
model (see Sec.~\ref{section:fid}).

We note that the overall behavior described in the results
(Sec.~\ref{section:results}) is robust. We keep in mind that our
neutrino treatment is simplified and that a complete Boltzmann solver
may give different quantitative values.

%%%%%%%%%%%%%%%%%%%%%%%%%%%%%%%%%%%%%%%%%%%%%%%%%%%%%%%%

\section{Results}
\label{section:results}

We use either progenitor u40 or WWs40 (see
Sec.~\ref{section:initialmodels}), and we employ the LS220 EoS, the
LS220+$\pi$ EoS or the LS220+$\Lambda$ EoS (see
Sec.~\ref{section:eos}). We name our simulations with the name of the
EoS (ls, $\pi$, $\Lambda$), followed by that of the progenitor (u40 or
WW40). Names and results are summarized in
Tab.~\ref{table:prop}. Simulations are stopped after the AH is formed.

\subsection{Fiducial model}
\label{section:fid}

We start by describing the simulation lsu40, using the LS220 EoS
with no extra particles and the u40 progenitor from \cite{woos_02}. This
simulation shall serve as a reference run in order to compare with simulations
using other EoSs and/or the other progenitor.

Both the collapse and bounce phases are followed using the leakage
scheme, which is possible because we solve advection equations for
$Y_{\nu_e}$ and $Y_{\bar{\nu}_e}$ as described in
Sec.~\ref{section:thequil}. The bounce is detected at the moment of
the shock formation, but, as there is no switch between the different
schemes before and after bounce, the exact moment of bounce detection
has very little influence on the dynamics (see \cite{sche_10} for an
example of an implementation using a switch at bounce).

After the bounce, the shock is launched and it reaches a distance of
about $98$ km from the center before it stalls. Then, the cooling of
the material makes the shock slowly recede. At the end of the
simulation, before the BH collapse is triggered, the shock remains
about $57$ km away from the center and its velocity has increased to
$0.35$c. The PNS radius, defined somewhat arbitrarily at the density of
$10^{11} \ \textrm{g.cm}^{-3}$ following \cite{muel_12b}, recedes,
too: At bounce, it is about $75$ km, then it shrinks slowly to about
$45$ km at the onset of the BH collapse.

\begin{figure*}
  \centering
  \includegraphics[scale=1.2]{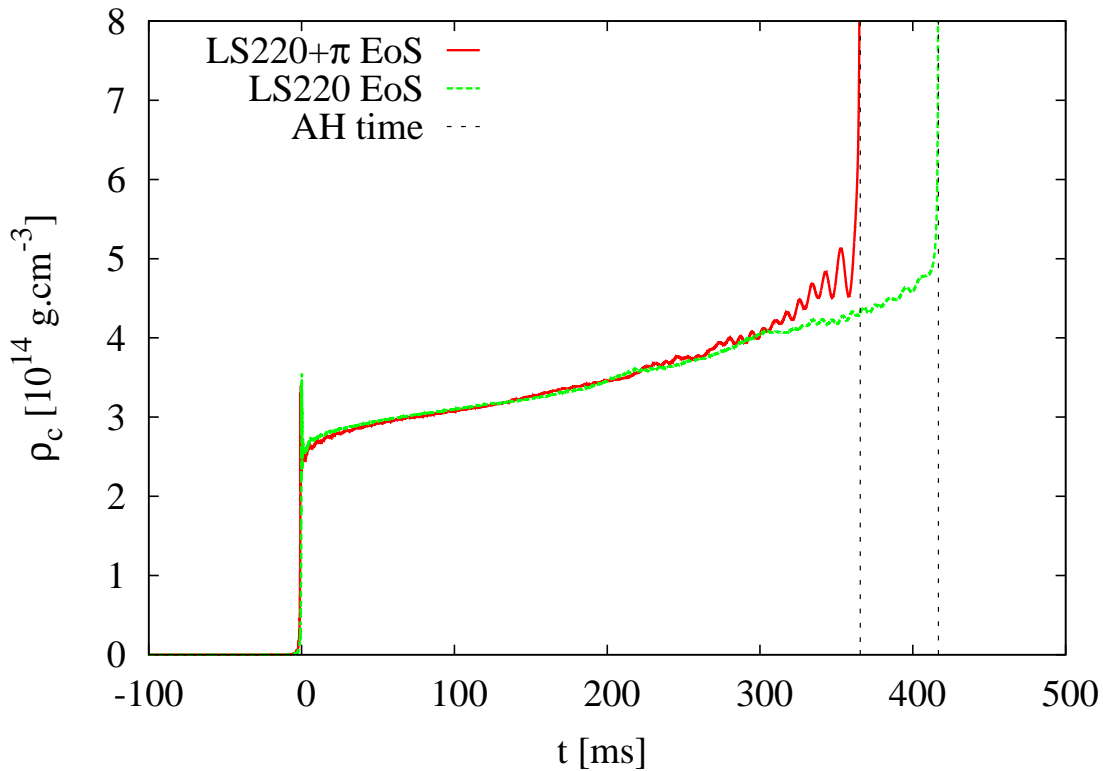}    
  \caption{Central density as a function of time post-bounce, for
    models lsu40 and $\pi$u40. The time corresponding to the first
    detection of the AH is marked by a vertical line.}
  \label{fig:rho_c_t_2}
\end{figure*}

As depicted in Fig.~\ref{fig:rho_c_t_2}, the PNS collapses at
$415$~ms post-bounce, within about 2~ms. The central density increases
from $4.6 \times 10^{14} \ \textrm{g.cm}^{-3}$ to $1.5 \times 10^{15}
\ \textrm{g.cm}^{-3}$ and the temperature raises from $4.3 \times
10^{11}$~K to $1.3 \times 10^{12}$~K. At 416.8~ms post bounce, the
apparent horizon is found and the black hole is formed.

The enclosed mass at the onset of BH collapse, measured by evaluating the 
baryonic mass up to the position of the shock, is $2.55 \
\textrm{M}_\odot$ (all masses given in this section are baryonic masses). It is a
measure (slightly overestimated, because the shock is not exactly at
the PNS radius) of the PNS maximum mass, above which it collapses to a
BH. This result is consistent with \cite{ocon_11} within $3\%$ (their
reported PNS mass is $2.469 \ \textrm{M}_\odot$).

\subsection{Model with pions and the u40 progenitor}

Adding pions to the EoS slightly softens it, so that we expect the PNS maximum
mass and the PNS radius to be slightly smaller.

During the entire simulation, $\pi^-$ are the most abundant
pions. This is to be expected since negatively charged particles are
favored within the neutron rich hot and dense matter encountered
during core-collapse. At bounce the maximum of $Y_{\pi^-}$ is
$Y_{\pi^-} = 0.0032$. The influence of pions at this moment becomes
non-negligible.

In Fig.~\ref{fig:rho_c_t_2}, we can see a consequence of the presence of pions
on the collapse phase, namely, the central density at bounce is slightly lower
in model $\pi$u40 than in model lsu40. Indeed, collapse and deleptonization
are slightly different, and it results in an homologous core and a nascent PNS
larger in model $\pi$u40. More precisely, the central temperature is $\sim
2\%$ higher at bounce and the difference stays relatively constant during the
first few tens of milliseconds, and this results in a $\sim 3\%$ lower central
density for model $\pi$u40 compared to model lsu40.

As can be seen in Fig.~\ref{fig:rho_c_t_2}, the PNS for $\pi$u40 is slightly
more compressible and the central density increases more rapidly. At $\sim
200$~ms post-bounce, there is a crossing point where the central densities of
both models are the same, then the model with pions predicts a higher density
until the collapse to the BH. The addition of pions only slightly changes the
behavior after bounce.

The central density in Fig.~\ref{fig:rho_c_t_2} shows oscillations as
it approaches the BH collapse time. We note that these oscillations
are more pronounced in model $\pi$u40 than in model lsu40. This might
be linked to the higher compressibility of the EoS due to the presence
of pions. These oscillations can be damped by medium effects or by
multidimensional effects (turbulence, \dots), and they do not change
the qualitative behavior of our simulations, as can be inferred from
Fig.~\ref{fig:rho_c_t_2}. So, we think they are not very important for
our study.

\begin{figure}
  \centering
  \includegraphics[width = .45\textwidth]{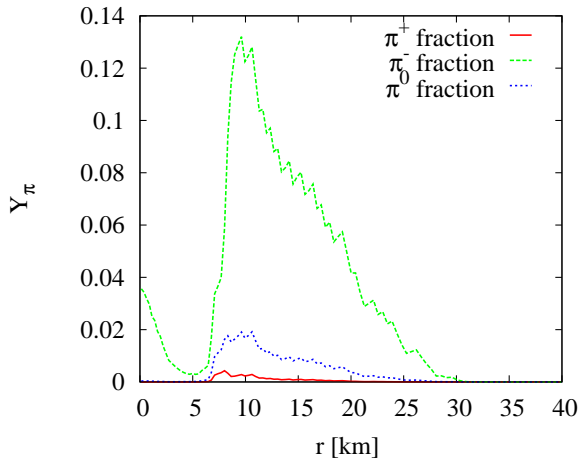}  
  \caption{Pion fractions as functions of radius at the onset of BH collapse
    ($t = 356 \ \textrm{ms}$ post-bounce, model $\pi$u40). The shape of the
    profile, in particular the decrease in the central region, is directly
    correlated with the temperature profile.}
  \label{fig:ypion}
\end{figure}

Fig.~\ref{fig:ypion} shows the pion fraction just before the collapse to the
BH at $t = 356$~ms post-bounce. As mentioned above, we observe that $\pi^-$
are the most abundant, followed by $\pi^0$ and then $\pi^+$.

We note that $\pi^-$ is the only species with a fraction greater than $2\%$,
and all species are present only in the very center of the PNS, up to $\sim
30$~km. The shape of the profile, i.e. the fact that near the center the pion
fractions are smaller, can be explained by the lower temperature near the
center, too. Pions are mainly thermally excited, so that their fractions
naturally follow closely the temperature profile. The pion fractions are of
the same order as those reported by~\cite{naka_12} using an extended
HShen EoS. The most abundant species, $\pi^-$, has a maximum of
$Y_{\pi^-} \simeq 0.13$ at the onset of the BH collapse. During a
simulation, all pion fractions increase, from 0 to their maximum value
at the BH collapse, following the increase of the temperature and the
density inside the PNS.

We find that the apparent horizon is detected at 366.3~ms post bounce
for a PNS maximum mass of $2.49 \ \textrm{M}_\odot$ with a PNS radius
of 43 km. We note that even though pion fractions are not very large
($Y_{\pi^-}$ is the only one over $0.10$, and only in the last tens of
milliseconds before BH collapse), this small change in the EoS results
in non-negligible changes in the dynamics, especially the time before
BH collapse.

\subsection{Model with $\Lambda$-hyperons and the u40 progenitor}

\subsubsection{Contraction of the PNS}
\label{sect:contractlambda}

\begin{figure*}
  \centering
  \includegraphics[scale=1.2]{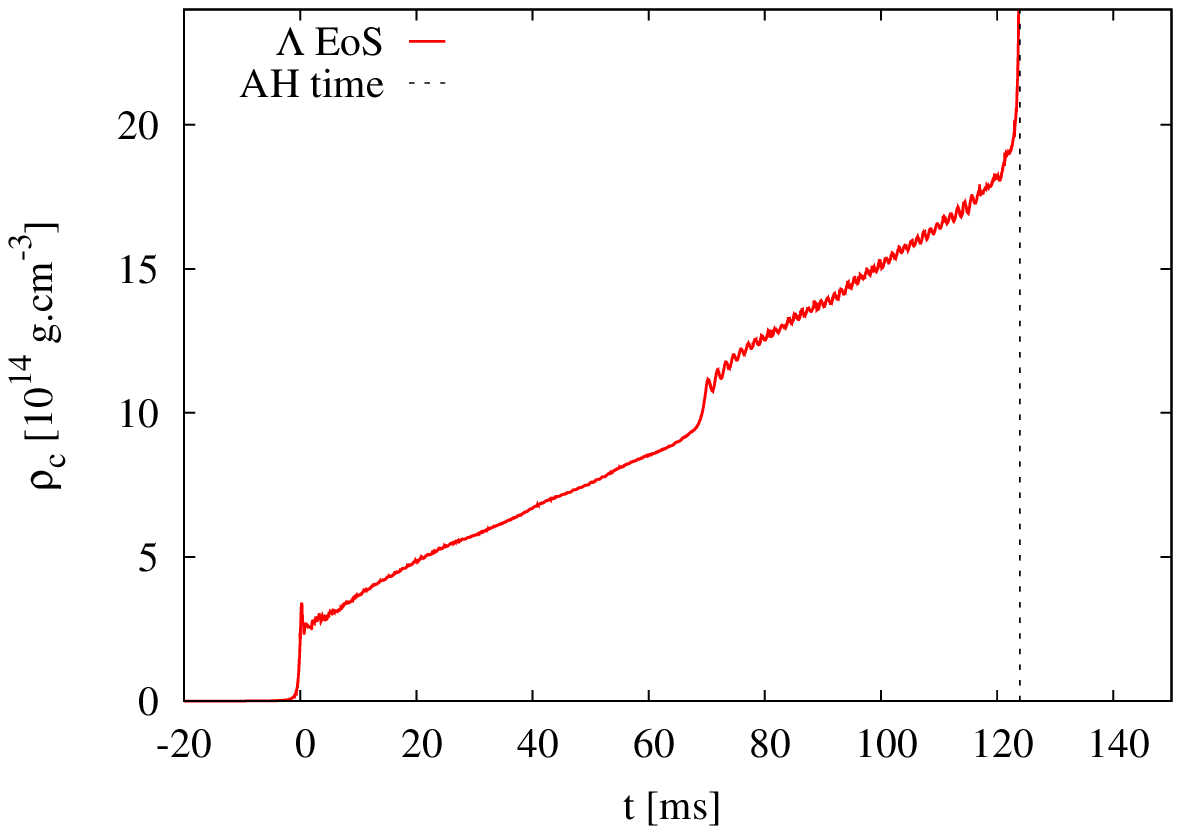}      
  \caption{Central density as a function of time post-bounce, for
    model $\Lambda$u40. The time corresponding to the first detection of
    the AH is marked by a vertical line.}
  \label{fig:rho_c_t_lambda}  
\end{figure*}

Adding $\Lambda$-hyperons to the EoS results in a more compact PNS, and
the effect is much more pronounced than with pions.

Indeed, in model $\Lambda$u40, the onset of collapse to BH occurs at
$122.8$~ms post-bounce and the AH is detected at $123.9$~ms
post-bounce, for a maximum mass of the PNS of only $2.00 \
\textrm{M}_\odot$. Note that this value is significantly lower than
its cold EoS counterpart (LS220+$\Lambda$ allows for cold neutron stars
with a maximum baryonic mass of $2.22 \ \textrm{M}_\odot$). This
difference might be attributed to dynamical effects. In addition, the hot
proto-neutron star is not entirely in $\beta$-equilibrium, such that the $Y_e$
profile is given dynamically. Since the EoS depends considerably on $Y_e$,
this might explain the lower supported mass, too. 

Fig.~\ref{fig:rho_c_t_lambda} shows that the central density increases
much more rapidly than in the fiducial model lsu40. The shock
propagates only up to $36$~km away from the center before
receding. The newly formed PNS extends to $31$~km a few milliseconds
after bounce and contracts to a radius of $17$~km very
rapidly. Contraction continues slowly all along the simulation up to
the phase transition (described in the next subsection), where the PNS
radius suddenly decreases from $15$~km to $13$~km. It further
contracts to about $10$~km at the onset of BH collapse. This radius is
very small for a hot PNS (see e.g. \cite{muel_12} for reported PNS
radii with another progenitor and the LS180 EoS. Radii typically lie
between $20$ and $70$ km).

\begin{figure}
  \centering
  \includegraphics[width = .45\textwidth,height = 6.25cm]{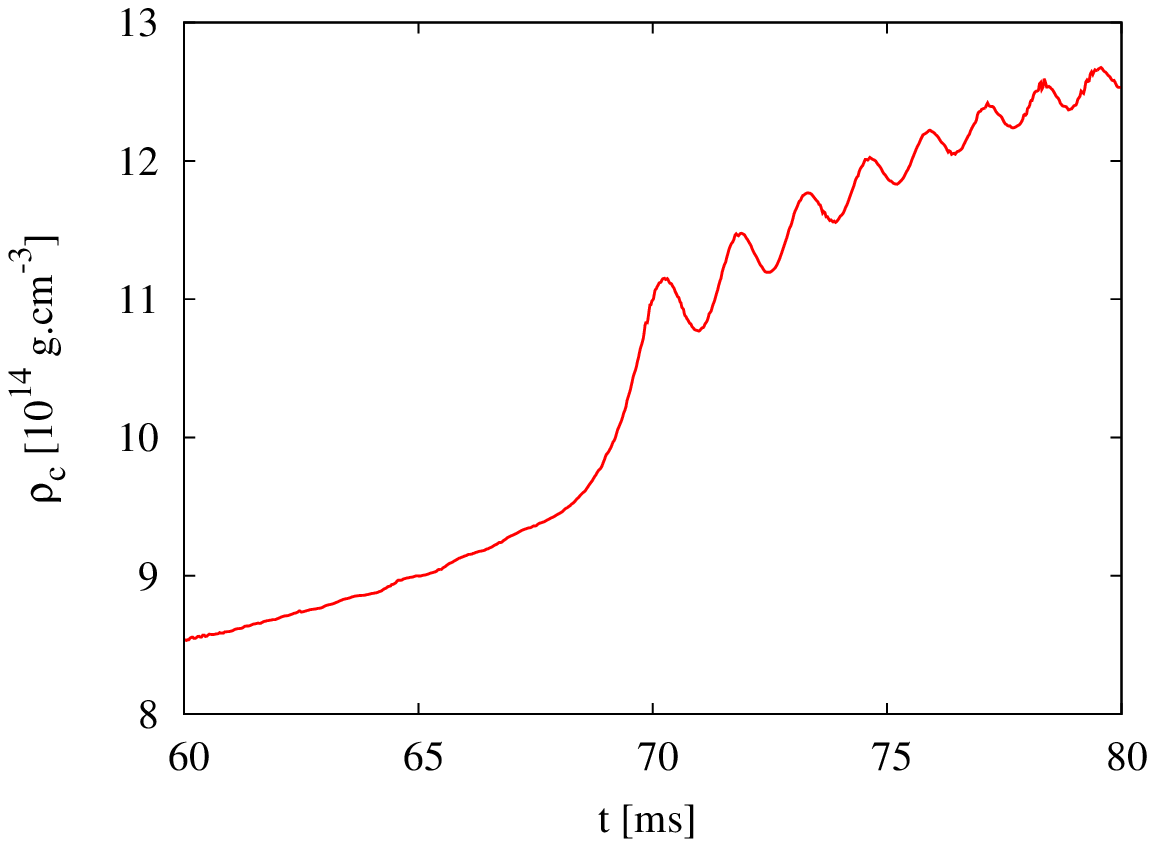}      
  \caption{Central density as a function of time post-bounce, model
    $\Lambda$u40, zoomed on the phase transition.}
  \label{fig:rho_c_t_lambda_zoom}
\end{figure}

\begin{figure}
  \centering
  \includegraphics[width = .45\textwidth]{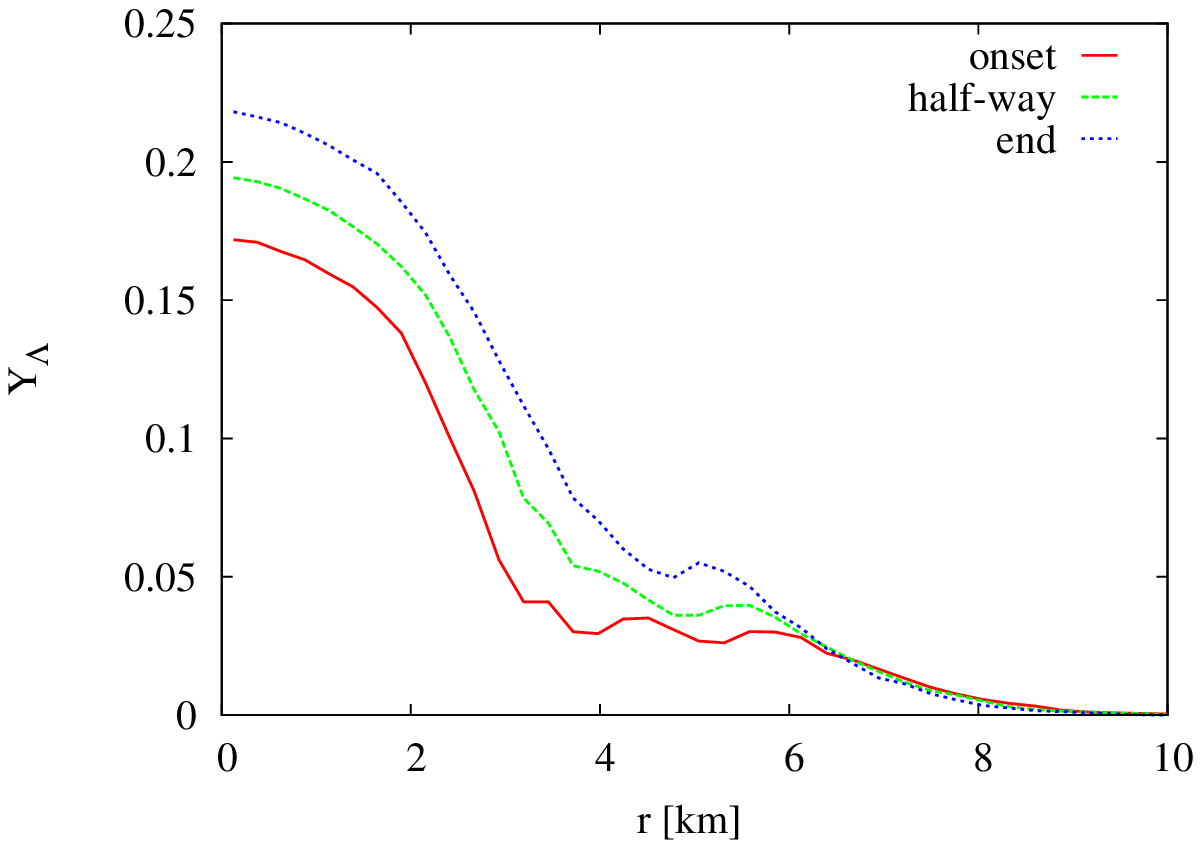}      
  \caption{$Y_\Lambda$ evolution during phase transition (model
    $\Lambda$u40). Snapshots are taken at the onset ($t = 68.7 \ \textrm{ms}$
    post-bounce), half-way through ($t = 69.5 \ \textrm{ms}$
    post-bounce) and at the end of the phase transition ($t = 70.2 \ \textrm{ms}$
    post-bounce).}
  \label{fig:y_lambda_phasetr}
\end{figure}

\subsubsection{Phase transition}
\label{section:phasetr}

The LS220+$\Lambda$ EoS contains a first order phase
transition to hyperonic matter, as discussed in
Sec.~\ref{section:eos}, which we expect to find in a core-collapse
simulation. Indeed, Fig.~\ref{fig:rho_c_t_lambda_zoom} is a detailed
view of Fig.~\ref{fig:rho_c_t_lambda}, where a phase transition
occurs. The density raises from $9.4 \times 10^{14} \
\textrm{g.cm}^{-3}$ to $11.2 \times 10^{14} \ \textrm{g.cm}^{-3}$
within less than $2$~ms (around 68-70~ms post-bounce).

This phase transition leads to a sudden contraction of the PNS on a
dynamical time scale, as discussed in
Sec.~\ref{sect:contractlambda}. Further contraction of the PNS is very
similar to the contraction of the iron core during the initial
collapse.

Fig.~\ref{fig:y_lambda_phasetr} shows the $\Lambda$-hyperon fraction
$Y_\Lambda$ as a function of radius during the phase transition. At
finite temperature, $\Lambda$-hyperons already appear at bounce, and
$Y_\Lambda$ keeps increasing to reach $\sim 0.17$ at the center at $68$~ms
post-bounce. Then, the phase transition occurs, and we can
see in Fig.~\ref{fig:y_lambda_phasetr} that $Y_\Lambda$ increases
from $\sim 0.17$ to $\sim 0.22$ at the center. This happens within
less than $2$~ms. After the phase transition, $Y_\Lambda$ oscillates
with the density and increases again at BH collapse.

\begin{figure}
  \centering
  \includegraphics[width = .45\textwidth]{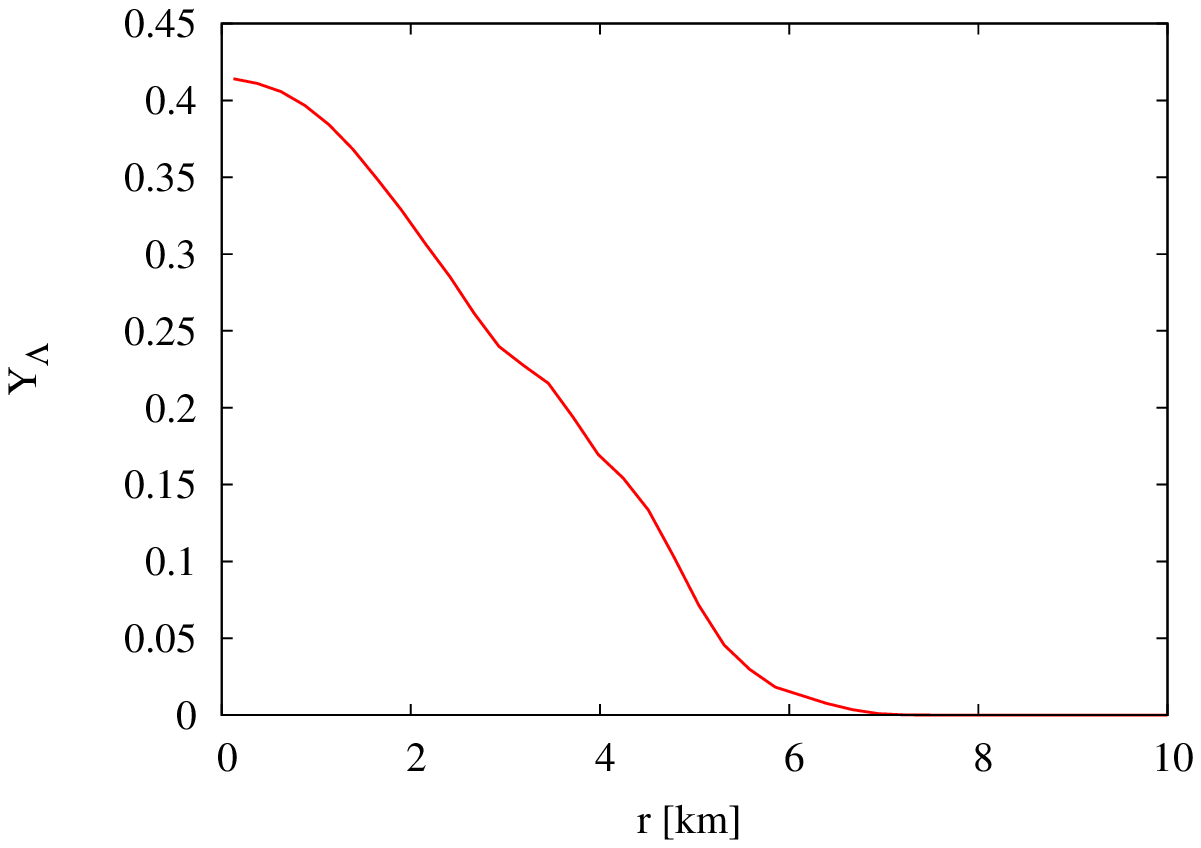}
  \caption{$Y_\Lambda$-profile (model $\Lambda$u40) at the onset of BH
    collapse ($t=121 \ \textrm{ms}$ post-bounce).}
  \label{fig:ylambda}
\end{figure}

Fig.~\ref{fig:ylambda} shows $Y_\Lambda$ at the onset of BH collapse. We see
that $Y_\Lambda$ is quite large and can go up to $0.41$ at the very center of
the PNS. Note that the phase transition is a consequence of the very large
accretion rate of the progenitor, and other progenitors can give a PNS that
collapses directly to BH before reaching the region of phase transition (see
Sec.~\ref{sect:ww40}).

Moreover, Fig.~\ref{fig:rho_c_t_lambda_zoom} shows oscillations of the
PNS excited by the collapse at phase transition. These could be
fundamental mode radial oscillations of the PNS (see \cite{kokk_01}
for a study in cold neutron stars, with different EoSs). It is not
possible to directly compare the oscillation frequency of our
simulations, which we estimate to be $\sim 800 \ \textrm{Hz}$, to the
data presented in \cite{kokk_01}, because i) the authors discuss only
cold neutron stars, ii) our system accretes matter continuously and
iii) we use none of the EoSs listed in their tables. Nevertheless, the
order of magnitude is correct and given the uncertainties on our
estimations of mass and radius of our PNS, we might interpret these
oscillations to be the fundamental radial mode of the PNS. In
addition, we performed several simulations with increased resolution
to rule out a numerical artifact, and we found these oscillations to
be robust.

\begin{table*}
  \caption{Properties at BH collapse.
    \label{table:prop}
  }
  \begin{center}
    \begin{tabular}{ccccccc}
      \hline \hline
      Name & lsu40 & $\pi$u40 & $\Lambda$u40 & lsWW40 & $\pi$WW40 &
      $\Lambda$WW40 \\
      \hline
      Initial model &  u40    & u40 & u40 & WWs40 & WWs40 & WWs40 \\
      \hline
      EoS                           & ~LS220~  & ~LS220+$\pi$~ & ~LS220+$\Lambda$~ 
      & ~LS220~ & ~LS220+$\pi$~ & ~LS220+$\Lambda$ \\
      \hline
      AH (ms post-bounce) ~ ~       &  416.8   & 366.3      & 123.9    
      &  1025.8  & 607.2      & 274.5  \\
      PNS mass ($\textrm{M}_\odot$)  &  2.55    & 2.49       & 2.00 
      &  2.73   & 2.40         & 2.00    \\
      PNS radius (km)                 & 45       & 43          & 10 
      &  51     & 38           & 18  \\ 
      \hline \hline
    \end{tabular}
  \end{center}
\end{table*}

\subsubsection{Influence of the $\Lambda$-scattering}
\label{section:inflscatt}

To test the influence of our newly implemented isoenergetic scattering off
$\Lambda$ (see Sec.~\ref{section:nusph} and App.~\ref{app:microphys}), another
simulation is conducted using the same parameters and the same EoS, but
switching off the $\Lambda$-scattering, and we compare it to the simulation
with $\Lambda$-scattering.

Differences are negligible up to the phase transition. Then, without
scattering off $\Lambda$, the phase transition occurs $6$~ms later, and the
PNS maximum mass becomes $1.96 \ \textrm{M}_\odot$. This is a reduction of
$0.04 \ \textrm{M}_\odot$ compared with the case including
$\Lambda$-scattering. This reduction of the maximum mass suggests that cooling
is more important when we reduce the opacity by neglecting scattering off
$\Lambda$-hyperons, as expected. However, the reduction is within the error
bars of the code, so this conclusion should be taken with some care. We note
that although $\Lambda$-hyperons are abundant, they appear in a medium that is
already optically thick. This could explain why the $\Lambda$-scattering is
not very important.

\subsubsection{Neutrino luminosity}

As a complementary tool, we compute the total luminosity $L_\nu$. In spherical
symmetry, we approximate it by 

\begin{equation}
  \label{eq:lum}
  L_\nu = 4 \pi \int_0^\infty W \left( 1 + \frac{v_r}{\Phi^2} \right)
  \left[ \frac{1}{\alpha} - \frac{\beta^r}{\alpha} \right] Q_E
  \, \sqrt{-g}  \, \mathrm{d}r.
\end{equation}

$Q_E$ being defined in the fluid rest frame, we have to transform it
to the Eulerian frame with a Lorentz boost (with a Eulerian velocity
$v_r/\Phi^2$, see Appendix~\ref{app:hydro_sources}), and then to the
coordinate frame, which is done by multiplying by the terms in
brackets. The generalization to 3D would involve a Lorentz boost in an
arbitrary direction and the complete shift vector.

The neutrino luminosity can be found by integrating the energy rate on
an invariant volume element at constant time, which justifies the
$\sqrt{-g}$ term. This formula agrees with \cite{ocon_10,
  kuro_12}. Note that the integration at constant time is an
approximation, since a time delay between the different emitting
regions should exist. But because the overall emitting region is
narrow ($\sim 100$ km at most), this approximation should be well justified.

As already reported by different authors (e.g., \cite{kuro_12, naka_12}), the
total neutrino luminosity peaks at the neutrino burst, a few milliseconds
after the bounce, and decreases very rapidly afterwards. We observe the same
behavior (see also Fig.~\ref{fig:l_nu} and the discussion for the WWs40
progenitor in Sec.~\ref{sect:ww40}), and the luminosity for models lsu40 and
$\pi$u40 are very similar. Only $\Lambda$u40 shows a different trend.

The neutrino burst peak luminosity is higher, and because the PNS shrinks very
rapidly, high neutrino emission after bounce is more sustained with the model
$\Lambda$u40. As a result, the luminosity decrease after the neutrino burst is
slower in this model.

Moreover, after the phase transition the luminosity follows the PNS
behavior (illustrated by the central density plot in
Fig.~\ref{fig:rho_c_t_lambda}), and oscillates, which makes the PNS
loose a lot of energy. The differences in luminosity between LS220 EoS
and $\Lambda$EoS have to be checked with a better neutrino treatment
to infer the detectability of the LS220 + $\Lambda$ EoS peculiarities. In
addition, the detection of PNS modes could be linked to a physical
process different from the phase transition, and it would be hard to
disentangle on a neutrino light curve.

Finally, at BH formation, the emitting region enters the horizon and
the luminosity drops down abruptly.

\subsection{Results with a WWs40 progenitor}
\label{sect:ww40}

Within this section we discuss simulations performed using a $40 \ \textrm{M}_\odot$,
solar metalicity progenitor from \cite{woos_95}, varying between EoSs
LS220, LS220+$\pi$ and LS220+$\Lambda$. This progenitor has often been
used to study BH collapse \cite{ocon_11, fish_09, sumi_07, hemp_12},
as its iron core is reported to be very large. The main difference
with the u40 progenitor from \cite{woos_02} is that the accretion rate
is significantly higher in the latter.

\begin{figure*}
  \centering
  \includegraphics[scale=1.2]{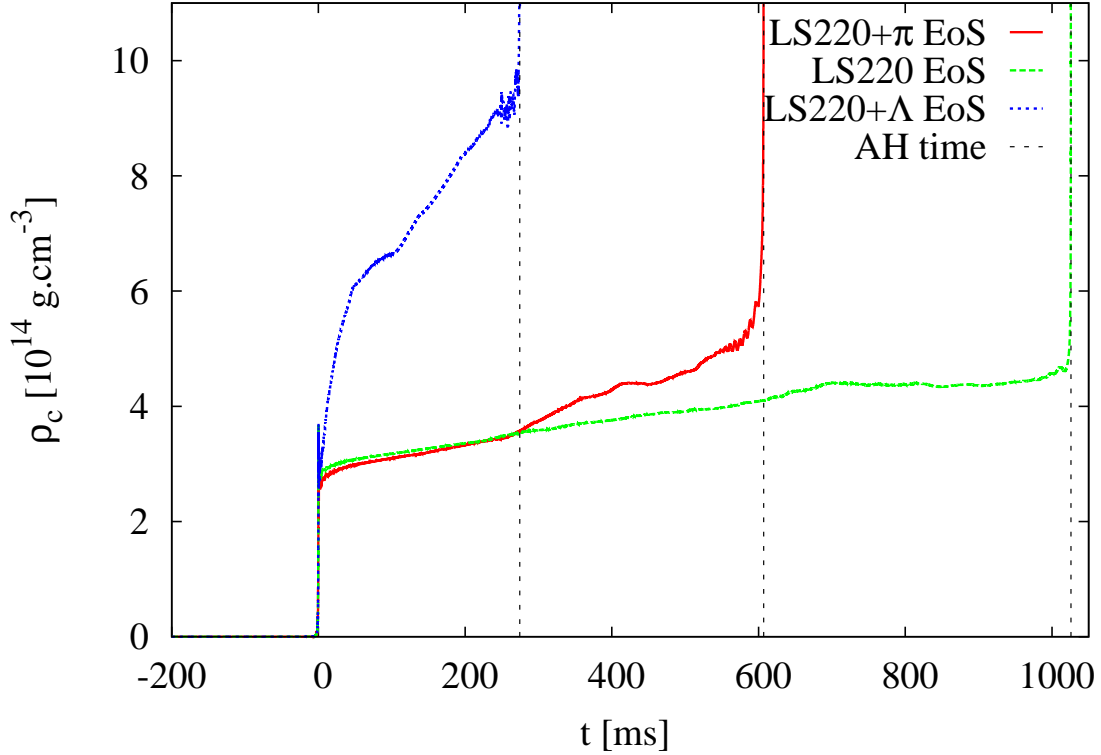}
  \caption{Central density as a function of time for model lsWW40
    (dashed, green), $\pi$WW40 (plain, red) and $\Lambda$WW40
    (dotted, blue).}
  \label{fig:rho_c_t_WW40}
\end{figure*}

\begin{figure}
  \centering
    \includegraphics[width=0.45\textwidth]{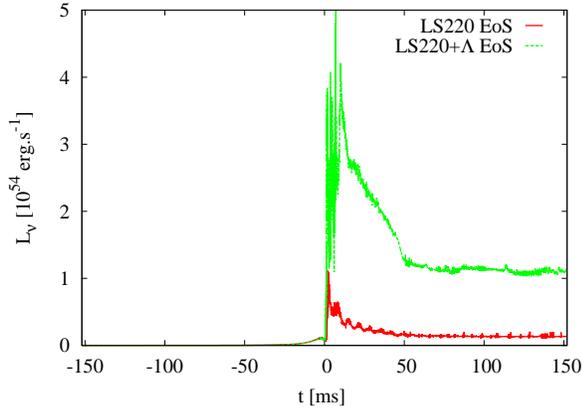}
    \caption{Neutrino luminosity as a function of post-bounce time for model
      $\Lambda$WW40 (dashed, green), compared to reference model lsWW40
      (plain, red). Model $\Lambda$WW40 shows a very high peak luminosity and
      a more sustained neutrino burst.}
  \label{fig:l_nu}
\end{figure}

In Fig.~\ref{fig:rho_c_t_WW40}, we observe the same trend as for progenitor
u40. With the model $\pi$WW40, the bounce occurs at lower density than with
the model lsWW40, and after $275$~ms the curves cross, due to the fact that
LS220+$\pi$ EoS allows the PNS to contract faster. For model lsWWs40 we find a
maximum PNS mass of $2.73 \ \textrm{M}_\odot$. Compared with previous
work~\cite{ocon_11}, this value agrees within about $12\%$, and may be
slightly overestimated. Other authors \cite{sumi_07, fish_09} often use a
lower value of the coefficient of nuclear incompressibility, namely $K = 180 \
\textrm{MeV}$, while we use $K = 220 \ \textrm{MeV}$. This significantly
changes the maximum mass that the PNS can hold and thus, comparisons are more
difficult.

The model lsWW40 forms an AH at $1025.8$~ms post-bounce, while the model
$\pi$WW40 collapses earlier, and an AH is detected at $607.2$~ms, for a
maximum PNS mass of $2.40 \ \textrm{M}_\odot$. We note that the addition of
pions leads to larger changes in the properties at BH collapse for WWs40 than
for u40 (see Tab.~\ref{table:prop}). We may interpret this as follows: because
the PNS contracts more when using LS220+$\pi$ EoS, the neutrinosphere is
closer to the center and so, cooling is more effective. Because accretion rate
is lower with the WWs40 progenitor, more time is needed to reach the PNS
maximum mass. Hence, the more effective cooling lasts longer and differences
between models using LS220+$\pi$ EoS and LS220 EoS are more pronounced.

Fig.~\ref{fig:rho_c_t_WW40} has been translated with the respective
bounce time of each model. Note that the bounce happens $\sim 12$~ms
earlier for model $\Lambda$WW40, as a small fraction of $\Lambda$
begins to appear ($Y_\Lambda \sim 0.0005$ at the center at bounce)
before bounce. For this model BH collapse is triggered before
the phase transition density is reached. Indeed, with progenitor u40
we reach a central 
density of $10^{15} \ \textrm{g.cm}^{-3}$ at $\sim 69$~ms post-bounce,
which triggers the onset of the phase transition. Here, as can be seen in
Fig.~\ref{fig:rho_c_t_WW40}, the collapse to the BH has already
started when we reach a density of $10^{15} \ \textrm{g.cm}^{-3}$. This is due to
the higher accretion rate of progenitor u40, which allows for higher
central densities of the PNS. Hence, the LS220+$\Lambda$ EoS does not
induce a phase transition for every progenitor that collapses to a
BH. Note that, even without phase transition, we find a non-negligible
fraction of $\Lambda$-hyperons ($Y_\Lambda \sim 0.15$ at the center at
the onset of BH collapse, at $t = 271 \ \textrm{ms}$ post-bounce. See
Fig.~\ref{fig:y_lambda_phasetr}, too, which shows $Y_\Lambda \sim
0.17$ at the center in model $\Lambda$u40 before the phase
transition). This is, as noted in Sec.~\ref{section:phasetr}, due
to the fact that $\Lambda$-hyperons begin to appear before the phase
transition at finite temperature, and their fraction increases with
increasing temperature. 

Finally, Fig.~\ref{fig:l_nu} shows the total neutrino luminosity $L_\nu$
(summed over all neutrino species) around the time of neutrino burst. Model
$\Lambda$WW40 shows a high peak luminosity of $5.10^{54} \
\textrm{erg.s}^{-1}$, compared to the reference model lsWW40 that shows a
maximum of $1.10^{54} \ \textrm{erg.s}^{-1}$. In the model $\Lambda$WW40, the
neutrino burst is also longer in time due to the rapid shrinking of the PNS,
and the post-bounce luminosity stays higher until BH formation.

%%%%%%%%%%%%%%%%%%%%%%%%%%%%%%%%%%%%%%%%%%%%%%%%%%%%%%%%

\section{Conclusion}

We have presented simulations of stellar core-collapses to BH, comparing
different finite-temperature EoSs with additional particles, namely pions and
$\Lambda$-hyperons. As expected, additional degrees of freedom modify the EoS
properties, in such a way that the collapse to a black hole occurs sooner
after the bounce.

Our new EoSs are based on the Lattimer and Swesty EoS. Pions are added as a
free gas, and $\Lambda$-hyperons are incorporated with the interactions
adapted from Balberg and Gal~\cite{balb_97}.  This EoS is subject to a
first-order phase transition driven by
$\Lambda$-hyperons~\cite{gulm_12a,gulm_12b} which is described by a Gibbs
construction. The LS220+$\pi$ EoS fulfills the gravitational mass constraint
from the $M = 1.97 \pm 0.04 \ \textrm{M}_\odot$ neutron star, and the
LS220+$\Lambda$ EoS is very slightly below with a maximum mass of $1.91 \
\textrm{M}_\odot$. Compared to previous works, H.Shen {\it et al.}
\cite{hshen_11} reported a maximum mass of $1.75 \ \textrm{M}_\odot$ for their
EoS including $\Lambda$-hyperons, and Ishizuka {\it et al.}  \cite{ishi_08}
reported a maximum mass of $1.55, 1.63$ and $1.65 \ \textrm{M}_\odot$ for
their EoSs that include different parameterizations of hyperons, muons and
pions. So, the EoSs presented in this paper are in better agreement with
observational mass constraints.

We have implemented a new leakage scheme. Compared with previous works, we
keep track of the neutrino fractions more accurately with advection equations,
and take into account neutrinos that are not trapped anymore because of the
neutrinosphere moving inwards. Although our fluid rest frame source terms are
approximated, we consistently transform them to the Eulerian frame, where the
hydrodynamic equations are solved. With these refinements, it is possible to
follow a simulation during the collapse, bounce and post-bounce phases without
changing the approach.

When using the LS220+$\Lambda$ EoS, we have implemented for the first time the
isoenergetic scattering off a $\Lambda$-hyperon. It is done in a simplified
manner consistent with our other opacity sources, and we find that it shifts
the occurrence of the phase transition by $6$ ms and the PNS maximum baryonic
mass is $0.04 \ \textrm{M}_\odot$ smaller with this opacity taken into
account. This contribution is still very subdominant compared with scattering
off nucleons.

Our use of the leakage scheme during the collapse phase enabled us to
see differences in the deleptonization, which resulted in a smaller
density at bounce for model $\pi$u40 and $\pi$WW40, compared to lsu40
and lsWW40, respectively. With both progenitors, we find, as expected,
that the compression of the PNS is faster when using the LS220+$\pi$ EoS. The
reduction of the PNS maximum baryonic mass is $0.06 \
\textrm{M}_\odot$ for model $\pi$u40 compared to model lsu40. A larger
difference of $0.33 \ \textrm{M}_\odot$ is found when comparing models
$\pi$WW40 and lsWW40. We interpret the larger difference as a result
of the larger amount of time spent in the post-bounce cooling
regime. Small differences between EoSs have thus more time to develop.

With the u40 progenitor and the LS220+$\Lambda$ EoS, a phase
transition clearly appears. Contrary to the cold $\Lambda$ EoS, at
finite temperature $\Lambda$-hyperons begin to appear before the phase
transition, and the fraction of thermally populated $\Lambda$-hyperons
is significant ($\sim 0.17$ at the center before the phase
transition). The phase transition results in a sudden increase in
density, comparable to the free fall during the initial collapse.

In contrast to model $\Lambda$u40, with model $\Lambda$WW40 we do not
observe the phase transition. We find that up to the onset of BH
collapse, the density stays too low to trigger the phase
transition. Consequently, the appearance of $\Lambda$-hyperons is
entirely due to thermal effects. We conclude that our models with
LS220+$\Lambda$ EoS do not predict a phase transition for every
progenitor. Indeed, only the progenitors with the highest accretion
rates are able to reach the phase transition density before the onset
of BH collapse.

Apart from the lack of the phase transition, the WWs40 progenitor has the same
qualitative behavior as u40. Our results with model lsWW40 are in agreement
with previous work with the same progenitor and the same EoS \cite{ocon_11}.
The LS220+$\pi$ EoS collapses to BH earlier than the LS220 EoS, admitting a
lower PNS maximum mass. LS220+$\Lambda$ EoS also leads to a more compressible
PNS, and the effect is much more pronounced. The PNS maximum mass is found to
be $2.00 \ \textrm{M}_\odot$.

Thus, we addressed the question of additional particles in hot EoSs. It
seems that adding pions or $\Lambda$-hyperons can significantly change
the conditions of a core-collapse supernova, while still being in
agreement with the $M = 1.97 \pm 0.04 \ \textrm{M}_\odot$ neutron
star. Because we see differences already at bounce, and because the neutrino
luminosity is higher when adding $\Lambda$-hyperons, one could infer
that additional particles may make a difference in the explosion
phase. This will have to be investigated with a more detailed neutrino
treatment.

The present EoS with $\Lambda$-hyperons is only marginally compatible
with the mass of PSR J 1614-2230 and the question arises to which
extent our results would be modified by taking another EoS giving a
higher maximum neutron star mass. As far as the same model is used,
e.g. with one of the parameter sets from Ref.~\cite{oert_12}, no
qualitative changes are to be expected. Indeed, as can be seen from
Ref.~\cite{oert_12}, the behavior of the hyperonic EoS with different
parameterizations are very similar. More pronounced modifications are
of course to be expected if a future observation gives an even higher
neutron star mass and another model has to be used (remind that LS220
EoS without additional particles has a maximum mass of only 2.06
M$_\odot$). The general effect of reduced time to black hole collapse
in presence of additional particles seems, however, very robust, since
we confirm the results from~\cite{ishi_08,sumi_09,naka_12} which have
been obtained using an extended version of the HShen EoS, thus a
different model for dense matter.

Further studies on observational consequences of the appearance of hyperons or
pions are yet to be done. In particular, the phase transition should produce
copious amount of gravitational waves, as it has been shown by previous
studies on ``mini-collapses''~\cite{lin_06, dimm_09}. A future study shall use
simulations of rotating stellar core collapse with pion and hyperon EoS, in
order to infer the gravitational wave signal. On the other hand, our
estimation of neutrino luminosity suggests that the PNS radial modes might be
detectable, although it would be hard to unambiguously associate it to a phase
transition to hyperonic matter, also because other phase transitions (e.g. to
quark matter~\cite{sage_09}) are possible. So, it seems possible to detect
evidences of the phase transition in an ideal case, combining observations of
neutrinos and gravitational waves from core-collapse to a BH, even without
constraining the nature of the phase transition from these data.

% would give a strong evidence of a phase
%transition, although it seems for the moment difficult to detect

Our work on modern EoSs including pions and $\Lambda$-hyperons raises
the question of the impact of the other additional particles that we
did not consider. In the long run, it will be desirable to build an
EoS with $\pi$, $\Lambda$ but also other hyperons ($\Sigma$, $\Xi$),
and muons. This EoS would have to fulfill the mass constraint and
consistently take into account interactions between
particles. Coupling it to an accurate neutrino transport scheme,
taking into account neutrino reactions with the new particles is the
only way to have accurate quantitative results on the influence of
additional particles. In the mean time, modern nuclear and
astrophysical data should restrict more and more the set of compatible
parameters, which will help building a realistic EoS.

\acknowledgments We would like to thank A. Fantina, M. Liebend\"orfer,
A. Perego for useful discussions, A.J. Penner for a careful reading of
the manuscript, and A. Heger for providing us with the progenitor data
from~\cite{woos_02} and~\cite{woos_95}. This work has been partially
funded by the SN2NS project ANR-10-BLAN-0503 and it has been supported
by Compstar, a research networking program of the European Science
Foundation.

%%%%%%%%%%%%%%%%%%%%%%%%%%%%%%%%%%%%%%%%%%%%
\begin{appendix}
\section{Explicit derivation of the hydrodynamic sources}
\label{app:hydro_sources}

Neutrinos enter the fluid momentum and energy equations as a source term in
the hydrodynamic Eqs.~$\nabla_\beta T^{\alpha \beta} = q^\alpha$, see
Eqs.~(\ref{eq:tmunusource}, \ref{eq:qalpha}). The quantities in
Eq.~(\ref{eq:qalpha}) being defined in the Lagrangian frame (LF), we
further need to transform them in the Eulerian frame (EF), where the
hydrodynamic equations, Eqs.~(\ref{eq:hydro}), are solved.

We adopt the same conventions as \cite{3plus1} and \cite{muel_10} for
the definitions of frames : The coordinate frame (that can be
associated with the grid) is fixed, and its tetrad $\partial_\mu$ is
used to define the metric $g_{\mu \nu}$ as (see \cite{3plus1})

\begin{equation}
  \label{eq:gmunu}
  g_{\mu \nu} = \bm{g} (\partial_\mu, \partial_\nu).
\end{equation}

Note also that the coordinate frame is not associated with a physical
observer. The Eulerian frame is then defined as in
Sec.~\ref{subsection:grhydro} (the Eulerian observer moves orthogonally
to spacelike hypersurfaces). It is an inertial frame, in which the
basis vectors span the Minkowski metric $\eta_{\alpha \beta} =
\textrm{diag}(-1, 1, 1, 1)$, so that it
is only in the Newtonian limit that the Eulerian frame becomes the
same as the coordinate frame.

This justifies that the transformation from the LF to the EF is a Lorentz
boost and contains no metric terms. Note, however, that our velocity $v_i$
defined in Sec.~\ref{subsection:grhydro} is written in the coordinate frame,
and we need to transform it into the EF.

To do so, we define the matrix of a general Lorentz transformation (tetrad
transformation) $P_\mu\,^{\nu}$ as

\begin{equation}
  \label{eq:vierbein}
  g_{\mu \nu} P_\alpha\,^\mu P_\beta\,^\nu =
  \eta_{\alpha \beta}.
\end{equation}

Going from the coordinate frame velocity to the EF velocity is done by applying this
transformation matrix $P$ to the 4-vector $v^\mu = (0, v^i)$. In the CFC, this
simply results in multiplying $v_i$ by $\textrm{diag} (1/\Phi^2, 1/(\Phi^2 r),
1/(\Phi^2 r \sin \theta))$.

Writing the boost for an arbitrary Eulerian velocity $v_{i,EF} = \left(
  v_{r}, v_{\theta}, v_{\phi} \right)$ leads to

\begin{equation}
  \label{eq:boost}
  q^{\alpha}_{EF} = \Lambda^{\alpha}\,_{\beta}(v^{i,EF}) q^\beta_{LF}
\end{equation}

and explicitly, the covariant components are

\begin{equation}
  \label{eq:boost0}
  q_{0,EF} = W(Q_E + v^{i,EF} Q_{(M)i})
\end{equation}

\begin{eqnarray}
  \label{eq:boost1}
  q_{1,EF} =  v_{r} W Q_E &+& \left[1 + (W-1)\frac{v_{r}v^{r}}{v_iv^i}\right]Q_{(M)1} \nonumber\\
       &+& \left[(W-1)\frac{v_{r}v^{\theta}}{v_iv^i}\right]Q_{(M)2} \nonumber\\
       &+& \left[(W-1)\frac{v_{r}v^{\phi}}{v_iv^i}\right]Q_{(M)3}
\end{eqnarray}

\begin{eqnarray}
  \label{eq:boost2}
  q_{2} =  v_{\theta} W Q_E &+& \left[(W-1)\frac{v_{\theta}v^{r}}{v_iv^i}\right]Q_{(M)1} \nonumber\\
       &+& \left[1 + (W-1)\frac{v_{\theta}v^{\theta}}{v_iv^i}\right]Q_{(M)2} \nonumber\\
       &+& \left[(W-1)\frac{v_{\theta}v^{\phi}}{v_iv^i}\right]Q_{(M)3}
\end{eqnarray}

\begin{eqnarray}
  \label{eq:boost3}
  q_{3, EF} =  v_{\phi} W Q_E &+& \left[(W-1)\frac{v_{\phi}v^{r}}{v_iv^i}\right]Q_{(M)1} \nonumber\\
       &+& \left[(W-1)\frac{v_{\phi}v^{\theta}}{v_iv^i}\right]Q_{(M)2} \nonumber\\
       &+& \left[1 + (W-1)\frac{v_{\phi}v^{\phi}}{v_iv^i}\right]Q_{(M)3}
\end{eqnarray}

It can easily be seen that Eqs.~(\ref{eq:boost0}), (\ref{eq:boost1}),
(\ref{eq:boost2}), (\ref{eq:boost3}) reduce to the usual spherically
symmetric boost by imposing $v_\theta = v_\phi = 0$. Terms arising
from the deviation from spherical symmetry, while usually small, can
become non-negligible, for instance in the case of a rapidly rotating
core \cite{muel_10}.

Finally, in Eq.~(\ref{eq:hydro}), $q_{0, EF}$ enters as such in the
source term of the conserved quantity $\mathcal{E}$ corresponding to
the conservation of energy, and $q_{j, EF}$ enters as such in the
source term of the conserved quantity $S_j$ corresponding to the
conservation of momentum. Note that because of the definition of
Eq.~(\ref{eq:hydro}), we recover there a multiplicative metric term,
$\sqrt{-g}$.

\section{Neutrino reactions formulae}
\label{app:microphys}

In this appendix, we do not use $c = 1$ for clarity. The
temperature $T$ is in units of energy.

\subsection{Opacities}

\subsubsection{Elastic scattering off a proton}
The opacity $\tau_s(\nu p)$ is defined following Ruffert {\it et
  al.}~\cite{ruff_96}, with the cross section corresponding to the
transport cross section from Burrows {\it et al.}~\cite{burr_06}.

\begin{equation}
  \tau_s(\nu p) = C_{sp} n_p \sigma_0 \left( \frac{ T
  }{m_e c^2} \right)^2 Y_{NN} \frac{F_5(\eta_\nu)}{F_3(\eta_\nu)}~,
\end{equation}

where $n_p$ is the proton number density,  $T$ is the temperature,
$\eta_\nu$ is the degeneracy parameter, and $\sigma_0$ is the weak
interaction cross section, defined by

\begin{equation}
  \sigma_0 = \frac{4 (m_ec^2 G_F)^2}{\pi (\hbar c)^4}~,
\end{equation}

with $G_F$ being the Fermi constant, $m_e$ the electron mass, and
$F_p$ the Fermi integral of order $p$ defined by
Eq.~(\ref{eq:fermi_integral}). The $F_5 / F_3$ term comes from phase
space integration and corresponds to the energy opacities
in~\cite{ruff_96}, which are the ones we find in better agreement with
more detailed calculations. We also include the following
approximation for the nucleon-nucleon degeneracy factor $Y_{NN}$,

\begin{equation}
  Y_{NN} = \frac{Y_N}{1 + \frac{2}{3} \max (\frac{\mu_N}{T}, 0)}~,
\end{equation}

where $Y_N$ is the nucleon fraction, $\mu_N$ is the nucleon chemical
potential and $T$ is the temperature. Finally, the constant term
$C_{sp}$ takes the value

\begin{equation}
  C_{sp} = \frac{1}{24} \left[ 4 (C_V - 1)^2 + 5 \left(
      \frac{g_A}{g_V} \right)^2 \right]~,
\end{equation}

where $g_V$ and $g_A$ are the vector and axial coupling constants of
the weak interaction, and $C_V = \frac{1}{2} + 2 \sin^2 \theta_W$ with
$\theta_W$ the Weinberg angle.

\subsubsection{Elastic scattering off a neutron}
Following Ruffert {\it et al.}~\cite{ruff_96}, the opacity $\tau_s(\nu
n)$ for the elastic scattering off a neutron is

\begin{equation}
  \tau_s(\nu n) = C_{sn} n_n \sigma_0 \left( \frac{T
  }{m_e c^2} \right)^2 Y_{NN} \frac{F_5(\eta_\nu)}{F_3(\eta_\nu)}~,
\end{equation}

where $n_n$ is the neutron number density. The constant $C_{sn}$ takes
the value

\begin{equation}
  C_{sn} = \frac{1}{24} \left[ 1 + 5 \left( \frac{g_A}{g_V} \right)^2 \right]~.
\end{equation}

\subsubsection{Elastic scattering off a nucleus}
We follow Rampp and Janka \cite{ramp_02} here, and define the
opacity $\tau_s(\nu A)$ as

\begin{eqnarray}
  \tau_s(\nu A) &=& \frac{\sigma_0 A^2 T^2
  }{16 m_e c^2} n_A \frac{F_5(\eta_\nu)}{F_3(\eta_\nu)}\nonumber\\
  &\times& \left( (C_A - C_V) + 
    (2 - C_A - C_V) \frac{2Z - A}{A} \right)^2 \nonumber\\
  &\times& \frac{y_b - 1 + (1 +
    y_b) \textrm{e}^{-2y_b}}{y_b^2} \langle S(\epsilon) \rangle_{ion}~,
\end{eqnarray}

where $n_A$ is the heavy nuclei number density, and $C_A =
\frac{1}{2}$. The ion screening term $\langle S(\epsilon)
\rangle_{ion}$ is taken into account as in Horowitz \cite{horo_97} and
$y_b$ is defined as in Bruenn and Mezzacappa \cite{brue_97} and Rampp
and Janka \cite{ramp_02}, explicitely

\begin{equation}
  y_b = \frac{2}{5 (\hbar c)^2} (1.07 A)^{2/3} \langle \epsilon_\nu
      \rangle^2~,
\end{equation}

with $A$ the mass number of the average heavy nucleus, and $\langle
\epsilon_\nu \rangle$ the mean neutrino energy, defined by Eq.~\ref{eq:epsnu}.

\subsubsection{Absorption of a $\nu_e$ by a neutron}
The opacity $\tau_a(\nu_e n)$, from \cite{ruff_96}, for this reaction is

\begin{eqnarray}
  \tau_a(\nu_e n) &=& \frac{n_n}{4} \left[ 1 + 3 \left( \frac{g_A}{g_V}
    \right)^2 \right] \sigma_0 |V_{ud}|^2 \left( \frac{T}{m_ec^2} \right)^2 \nonumber\\
  &\times& \frac{F_5(\eta_{\nu_e})}{F_3(\eta_{\nu_e})}
  \left[ 1 + \exp \left( \eta_e -
      \frac{F_5(\eta_{\nu_e})}{F_4(\eta_{\nu_e})} \right) \right]^{-1}
\end{eqnarray}

with $\eta_e$ the degeneracy parameter of the electrons and $|V_{ud}|$
the CKM matrix element.

\subsubsection{Absorption of a $\bar{\nu}_e$ by a proton}
The opacity $\tau_a(\bar{\nu}_e p)$, from \cite{ruff_96}, for this reaction is
\begin{eqnarray}
  \tau_a(\bar{\nu}_e p) &=& \frac{n_p}{4} \left[ 1 + 3 \left( \frac{g_A}{g_V}
    \right)^2 \right] \sigma_0 |V_{ud}|^2\left( \frac{T}{m_ec^2} \right)^2 \nonumber\\
  &\times& \frac{F_5(\eta_{\bar{\nu}_e})}{F_3(\eta_{\bar{\nu}_e})}
  \left[ 1 + \exp \left( - \eta_e -
      \frac{F_5(\eta_{\bar{\nu}_e})}{F_4(\eta_{\bar{\nu}_e})} \right) \right]^{-1}
\end{eqnarray}

\subsubsection{Elastic scattering off a $\Lambda$-hyperon}

Following the same approximations as for the transport cross section
for coherent scattering off neutrons in~\cite{brue_85, ruff_96,
  burr_06} we obtain,
  \begin{equation}
    \label{eq:sigmal}
    \sigma_\Lambda = \frac{\sigma_0}{4} \left( \frac{1 + 5 c_{A_\Lambda}^2}{6} \right)~.
  \end{equation}
  The constant $c_{A_\Lambda}$ is the $\Lambda$ axial coupling
  constant to the neutral current and is assumed to take its value
  given by flavor $SU(3)$ symmetry, $c_{A_\Lambda} =
  -0.73$~\cite{Sava_97, Redd_98}\footnote{Note that the strangeness
    content $\Delta s$ of the nucleon has been neglected when deriving
    this value, see~\cite{Sava_97}. The value of $\Delta s\sim -0.1$
    is subject to very large uncertainties and taking it into account
    would only very slightly modify our results.}. 

Explicitely, the opacity $\tau_s(\nu \Lambda)$ for this reaction is

\begin{equation}
  \tau_s(\nu \Lambda) = n_{\Lambda} \sigma_\Lambda \left( \frac{T}{m_e
      c^2} \right)^2\frac{F_5(\eta_\nu)}{F_3(\eta_\nu)}~,
\end{equation}

where $n_{\Lambda}$ is the $\Lambda$-hyperon number density. Note that
we do not take into account a degeneracy factor in this case.

\subsection{Neutrino creation terms}

\subsubsection{Electron and positron captures}
To take into account electron and positron captures, we integrate the
following expression (see \cite{brue_85})

\begin{equation}
  \label{eq:integcapt}
  \Sigma = \frac{c}{n_b} \frac{4 \pi}{(hc)^3} \int_0^\infty E^2 j (1 - f) \mathrm{d}E~,
\end{equation}

where $\Sigma = \Sigma_{ec}, \Sigma_{pc}$; $f$ is the neutrino
distribution function, which we assume is a Fermi-Dirac, because no
deviation from equilibrium is computed in the leakage scheme. Note
that we do not take into account the absorption term, and rather set
$\beta$-equilibrium at a given density.

These integrated rates are computationally expensive, and only require
the knowledge of the EoS to be computed. Therefore, they can be
tabulated as a function of $\rho, T, Y_e$ (at each point of the EoS),
and the effective neutrino chemical potential
$\mu_{\nu,\mathit{eff}}$. Access to the tables is done by a
quadrilinear interpolation.

$j$ is the creation term, also taken from Bruenn \cite{brue_85}.

\subsubsection{Electron capture on free protons}
The total electron capture rate is the sum of the electron capture
rate on free protons and the electron capture rate on nuclei.
For electron capture on free protons, $j_{ec,p}$ takes the value

\begin{eqnarray} 
  j_{ec,p} &=& \frac{G_F^2}{\pi (\hbar c)^4} |V_{ud}|^2 \eta_{pn} (g_V^2 + 3g_A^2)
  f_e(E+Q) \nonumber\\
  &\times& (E+Q)^2 \left[ 1 - \left( \frac{m_ec^2}{E + Q} \right)^2 \right]^{1/2}~,
\end{eqnarray}

where $Q$ is the mass difference between neutrons and protons, and
$\eta_{pn}$ is such that

\begin{equation}
  \eta_{pn} = \int 2 \frac{\mathrm{d}^3p}{(2 \pi)^3} f_n(E) (1 - f_p(E))~,
\end{equation}

with $f_e$, $f_n$ and $f_p$ the distribution functions of the
electrons, neutrons and protons, respectively, all taken to be
Fermi-Dirac distribution functions.

\subsubsection{Electron capture on nuclei}
For electron capture on nuclei, $j_{ec,n}$ takes the value

\begin{eqnarray}
  j_{ec,n} &=& \frac{G_F^2}{\pi (\hbar c)^4} n_A |V_{ud}|^2 g_A^2 N_p(Z)N_h(N) f_e(E+Q') \nonumber\\
  &\times& (E+Q')^2  \left[ 1 - \left( \frac{m_ec^2}{E + Q'} \right)^2 \right]^{1/2}~,
\end{eqnarray}

with the approximation $Q' = \mu_n - \mu_p + \Delta$, $\mu_n$ and
$\mu_p$ being the neutron and proton chemical potential, respectively,
and taking the constant value $\Delta = 3 \ \textrm{MeV}$.

$N_p(Z)$ and $N_h(N)$ are also taken following Bruenn \cite{brue_85},

\begin{equation}
  N_p(Z) = \left\{ \begin{array}{cc}
    0,  & Z < 20 \\
    Z-20, & 20 < Z < 28 \\
    8, & Z > 28 \end{array} \right.
\end{equation}

\begin{equation}
  N_h(N) = \left\{ \begin{array}{cc}
    6,  & N < 34 \\
    40-N, & 34 < N < 40 \\
    0, & N > 40 \end{array} \right.
\end{equation}

We start the integration Eq.~(\ref{eq:integcapt}) at the threshold $E = m_ec^2 -
Q'$ if this value is positive.

\subsubsection{Positron capture on free neutrons}
For positron capture on free neutrons, $j_{pc}$ takes the value

\begin{eqnarray}
  j_{pc} &=& \frac{G_F^2}{\pi (\hbar c)^4} |V_{ud}|^2 \eta_{np} (g_V^2 + 3g_A^2)
  f_{e^+}(E-Q) \nonumber\\
  &\times& (E-Q)^2 \left[ 1 - \left( \frac{m_ec^2}{E - Q}
    \right)^2  \right]^{1/2}~,
\end{eqnarray}

where $f_{e^+}$ is again a Fermi-Dirac distribution
function. $\eta_{np}$ can be inferred from the definition of
$\eta_{pn}$ by interchanging $n$ and $p$. We start the integration
Eq.~(\ref{eq:integcapt}) at the threshold $E = Q + m_ec^2$.

\subsubsection{Electron positron pair annihilation}
Following Ruffert {\it et al.}~\cite{ruff_96},

\begin{equation}
  \Sigma_{ee} = C_{ee} \frac{\sigma_0 c}{(m_ec^2)^2} \epsilon_{e^-}
  \epsilon_{e^+}  \langle 1 - f_{\nu_i} \rangle_{ee} \langle 1 -
  f_{\bar{\nu}_i} \rangle_{ee}~,
\end{equation}

with the definition

\begin{equation}
  \epsilon_{e^{\mp}} = \frac{8 \pi}{(hc)^3} T^4 F_3 (\pm \eta_e)~,
\end{equation}

and the average Fermi distributions

\begin{equation}
  \langle 1 - f_{\nu_i} \rangle_{ee} = \left( 1 + \exp \left[ - \left(
        \frac{F_4(\eta_e)}{2 F_3(\eta_e)} + \frac{F_4(- \eta_e)}{2
          F_3(- \eta_e)} - \eta_{\nu_i} \right) \right] \right)^{-1}
\end{equation}

The constant $C_{ee}$ takes the value $C_{ee} = ((C_V - C_A)^2 +
  (C_V + C_A)^2) /36$ for $\nu_e$ and $\bar{\nu}_e$, and $C_{ee} =
  ((C_V - C_A)^2 + (C_V + C_A - 2)^2) /9$ for $\nu_x$.

\subsubsection{Plasmon decay}
Following Ruffert {\it et al.}~\cite{ruff_96},

\begin{eqnarray}
  \Sigma_{pl} &=& \frac{\pi^2}{3\alpha_{em}} C_{pl}^2 \frac{\sigma_0
    c}{(m_ec^2)^2} \frac{T^8}{(hc)^6} \nonumber\\
  &\times& \gamma^6 \mathrm{e}^{-\gamma} (1
  + \gamma) \langle 1 - f_{\nu_i} \rangle_\gamma \langle 1 -
  f_{\bar{\nu}_i} \rangle_\gamma ~,
\end{eqnarray}

where $\alpha_{em}$ is the fine structure constant, $\gamma$ is
defined by

\begin{equation}
  \gamma = 2 \left( \sqrt{\frac{\alpha_{em}}{3 \pi}} \right)^{-1}
  \sqrt{\frac{1}{3}(\pi^2 + 3\eta_e^2)}~,
\end{equation}

and the average Fermi distributions are

\begin{equation}
  \langle 1 - f_{\nu_i} \rangle_\gamma = \left( 1 + \exp \left[ -
        \left( 1 + \frac{\gamma^2}{2 (1 + \gamma)} - \eta_{\nu_i}
        \right) \right] \right)^{-1}
\end{equation}

Note here that $\nu_x$ are taken into account with a degeneracy of 4
and a vanishing chemical potential. The constant $C_{pl}$ takes the
value $C_{pl} = C_V$ for $\nu_e$ and $\bar{\nu}_e$, and $C_{pl} = C_V
- 1$ for $\nu_x$.

\end{appendix}
%%%%%%%%%%%%%%%%%%%%%%%%%%%%%%%%%%%%%%%%%%%%%%%%%%%%%%%%

%\bibliography{biblio_sn}

\end{document}